\documentclass[aps,prd,preprint,nofootinbib,11pt]{revtex4}
\usepackage{amsfonts}
\usepackage{mathrsfs}
\usepackage{graphicx}
\usepackage{amsmath}
\usepackage{amssymb}
\usepackage{subfigure}
\usepackage{epsfig}
\usepackage{graphicx}
\newcommand{\bqa}{\begin{eqnarray}}
\newcommand{\eqa}{\end{eqnarray}}
\newcommand{\beq}{\begin{equation}}
\newcommand{\eeq}{\end{equation}}
\allowdisplaybreaks[1]
\graphicspath{{fig/}{dia/}} \DeclareGraphicsExtensions{.eps}
\begin{document}

\title{Doubly Heavy Tetraquarks in QCD Sum Rules\\[0.7cm]}

\author{Liang Tang$^{1}$, Bin-Dong Wan$^{2}$, Kim Maltman$^{3}\footnote{kmaltman@yorku.ca}$, Cong-Feng Qiao$^{2,4}\footnote{qiaocf@ucas.ac.cn}$}
\affiliation{$^1$ College of Physics, Hebei Normal University, Shijiazhuang 050024, China\\
$^2$ School of Physics, University of CAS, Beijing 100049, China\\
$^3$ Department of Mathematics \& Statistics, York University, Toronto, ON M3J 1P3, Canada
\\
$^4$ CAS Center for Excellence in Particle Physics, Beijing 100049, China
}

\author{~\\~\\}


\begin{abstract}
\vspace{0.3cm}
In the framework of QCD Sum Rules we investigate the $q q^\prime \bar{Q} \bar{Q}$ tetraquark structure with quantum number $J^P = 1^+$, which are embedded in two types of configurations, the $[8_c]_{q \bar{Q}} \otimes [8_c]_{q^\prime \bar{Q}}$ and $[1_c]_{q \bar{Q}} \otimes [1_c]_{q^\prime \bar{Q}}$ with $Q = b, c$, $q = u$, and $q^\prime = d, s$. Our finding confirms the Lattice QCD result that the bottom tetraquark states could exist and their masses are evaluated. In the calculation, the non-perturbative condensate contributions up to dimension eight in operator product expansion are considered, and those terms linear to the strange quark mass $m_s$ are kept. It is found that for octet-octet configuration the masses of potential tetraquark states are about $11.28$ GeV for the $ud\bar{b}\bar{b}$ system, and $11.31$ and $11.34$ GeV for the $us\bar{b}\bar{b}$ system, which are above the corresponding two-meson thresholds. For molecular configuration, the corresponding masses are found below the thresholds, that is $10.36$ GeV and $10.48$ GeV, respectively. The possible tetraquark decay channels are analyzed and the strong decay rates are evaluated. The mass dependence on the radiative correction and the condensates is estimated. Moreover, the doubly charmed tetraquark states are also considered.

\end{abstract}
 \maketitle
\newpage

\section{Introduction}

In the framework of quantum chromodynamics (QCD) and quark model, various of hadron structures beyond the normal meson and baryon are yet rarely explored, like multiquark state, hybrid, and glueball, etc, which are theoretically legitimate to exist but experimentally hard to identify. With the first observation of $X(3872)$ by Belle Collaboration in 2003 \cite{X3872}, a series of candidates of exotic hadron states were claimed to be measured. It is reasonable to think that we might be now at the dawn of the enormous discovery of novel hadronic structures. In the scope of exotic states, tetraquark state belongs to the simplest extension of the normal hadron.

Doubly heavy tetraquarks composed of two heavy antiquarks ($\bar{Q}\bar{Q}$) and two light quarks ($q q^\prime$), where $Q=b, c$ and $q= u$, $q^\prime = d, s$, are of peculiar interest in physics. They do not experience quark annihilation process in their decay modes. In the family of tetraquark, the existence of stable doubly heavy tetraquark was first suggested by Jaffe forty years ago \cite{Jaffe:1976ig}. Within the diquark-antidiquark model, the attraction force induced by the color Coulomb potential of two antiquarks in color $3_c$ configuration enables the existence of stable $qq\prime \bar{Q} \bar{Q}$ tetraquark states possible in the $m_Q \to \infty$ limit\cite{Quigg:2018eza}. The properties of such states were explored in depth in a variety of theories, such as the MIT bag model \cite{Carlson:1987hh}, color flux-tube model\cite{Deng:2018kly} ,chiral quark model \cite{Zhang:2007mu, Pepin:1996id}, constituent quark model \cite{Vijande:2006jf, Park:2018wjk}, chiral perturbation \cite{Manohar:1992nd}, and lattice QCD \cite{Bicudo:2015vta, Francis:2016hui, Leskovec:2019ioa}.

In past thirty years, the QCD Sum Rules technique \cite{Shifman}, as a model independent approach in the study of hadron physics, especially hadron spectrum, has achieved a great triumph, at least qualitatively. Employing the QCD Sum Rules, in recent years people have performed some remarkable researches on exotic hadrons, including the doubly heavy tetraquark states \cite{Navarra:2007yw, Du:2012wp, Dias:2011mi, Wang:2017uld}. In Ref.\cite{Navarra:2007yw}, Navarra, Nielsen and Lee worked with the current of axial heavy diquark and scalar light anti-diquark, and found there possibly exist the $J^P = 1^+$ $Q Q-\bar{u} \bar{d}$ states. Ref.\cite{Du:2012wp} calculated the diquark-antidiquark states with various $Q Q \bar{q} \bar{q}'$ configurations and  $J^P$ quantum numbers. Besides, Dias {\it et al.} \cite{Dias:2011mi} also calculated the tetraquark spectrum but with the $QQ\bar{q}\bar{q}$ being in molecular structure. Ref.\cite{Wang:2017uld} constructed certain (axial vector diquark)-(scalar antidiquark) type currents and investigated the axial vector doubly heavy tetraquark states.

In practice, as noted in Ref. \cite{Tang:2016pcf}, there may exist another type of tetraquark configuration of $(Q \bar{q}) (Q\bar{q'})$, where the diquarks are in color octet as in a hybrid. Though the color-octet diquark itself is loosely bounded relative to the color singlet ones as in the molecular state, the diquark-diquark confining interaction in the former case tends to be stronger than the latter, due to the direct QCD effect. In this work, we analyze in the framework of QCD Sum Rules whether there exists stable $J^P = 1^+$  tetraquark state with color-octet-octet configuration or not, and also calculate the doubly heavy molecular tetraquark state by means of currents given in Ref.\cite{Francis:2016hui}. The decay properties of the concerned tetraquark states will be evaluated as well.

The paper is organized as follows: after the introduction, we present some primary formulas of the QCD Sum Rules used in our calculation in Sec.II. The numerical analysis and results are given in Sec.III. The last part is left for conclusions and discussion of the results.

\section{Formalism}

The procedure of QCD Sum Rules calculation \cite{Shifman, Reinders:1984sr, Narison:1989aq, P.Col} starts with the two-point correlation function constructed by hadronic currents. For an axial-vector state considered in this work, the two-point correlation function takes the following form:
\begin{eqnarray}
\Pi_{\mu \nu}(q) &=& i \int d^4 x e^{i q \cdot x} \langle 0 | T \bigg{\{} j_\mu(x), j_\nu^\dagger(0) \bigg{\}} |0 \rangle \nonumber \\
&=& (q_\mu q_\nu -q^2 g_{\mu \nu}) \Pi(q^2)\; .
\end{eqnarray}
Here, $j_\mu(x)$ and $j_\nu(0)$ are the relevant hadronic currents with $J^P = 1^+$ and the Lorentz-structure of the right-hand side is dictated by the conservation of the hadronic current.

For doubly heavy tetraquark in color-octet-octet configuration with quantum number $J^P = 1^+$, the lowest order currents may be constructed exclusively as
\begin{eqnarray}
j^{A, 1^+}_\mu(x) \!\!\! &=& \!\!\! \bigg[i \, \bar{Q}^j(x) \gamma_5 (t^a)_{jk} q^k(x) \bigg] \bigg[\, \bar{Q}^m(x) \gamma_\mu (t^a)_{mn} q^{\prime \, n}(x) \bigg]\; \label{current-1}  , \\
j^{B, 1^+}_\mu(x) \!\!\! &=& \!\!\! \bigg[ \, \bar{Q}^j(x) \gamma_\mu (t^a)_{jk} q^k(x) \bigg] \bigg[i \, \bar{Q}^m(x) \gamma_5 (t^a)_{mn} q^{\prime \, n}(x) \bigg]\; \label{current-2}  , \\
j^{C, 1^+}_\mu(x) \!\!\! &=& \!\!\! \bigg[\, \bar{Q}^j(x) (t^a)_{jk} q^k(x) \bigg] \bigg[\, \bar{Q}^m(x) \gamma_\mu \gamma_5 (t^a)_{mn} q^{\prime \, n}(x) \bigg]\; \label{current-3}  , \\
j^{D, 1^+}_\mu(x) \!\!\! &=& \!\!\! \bigg[ \, \bar{Q}^j(x) \gamma_\mu \gamma_5 (t^a)_{jk} q^k(x) \bigg] \bigg[\, \bar{Q}^m(x) (t^a)_{mn} q^{\prime \, n}(x) \bigg]\; \label{current-4}\  ,
\end{eqnarray}
which are similar to the ones used in the investigation of light tetraquarks \cite{Latorre:1985uy, Narison:1986vw, Wang:2006ri, Wang:2015nwa}.
Here, $j$, $k$, $m$, and $n$ denote color indices, $t^a$ are $SU_c(3)$ color group generators, $Q$ represents the heavy quark $b$ or $c$, $q$ stands for the up quark $u$, and $q^\prime$ for the down quark $d$ or the strange quark $s$. Here, the superscripts $A$, $B$, $C$, and $D$ indicate the currents composed of a $0^-$ color-octet diquark and a $1^-$ color-octet diquark; a $1^-$ color-octet diquark and a $0^-$ color-octet diquark; a $0^+$ color-octet diquark and a $1^+$ color-octet diquark; and a $1^+$ color-octet diquark and a $0^+$ color-octet diquark, respectively.

For molecular structure, the current goes as \cite{Francis:2016hui}
\begin{eqnarray}
j^E_\mu(x) = \bigg[\bar{Q}_a\gamma_5 q_a(x)\bigg] \bigg[ \bar{Q}_b(x)\gamma_\mu q_b^{\prime}(x)\bigg]
- \bigg[\bar{Q}_a(x) \gamma_5 q_a^{\prime}(x)\bigg] \bigg[ \bar{Q}_b(x)\gamma_\mu q_b(x)\bigg] \, ,
\end{eqnarray}
where the subscripts $a$ and $b$ are color indices, and the superscript $E$ means that it is in the molecular configuration.

In order to calculate the spectral density of the operator product expansion (OPE) side, the light quark ($q=u$, $d$ or $s$) and heavy quark ($Q=c$ or $b$) full propagators $S^q_{i j}(x)$ and $S^Q_{i j}(p)$ are employed, say
\begin{eqnarray}
S^q_{j k}(x) \! \! & = & \! \! \frac{i \delta_{j k} x\!\!\!\slash}{2 \pi^2
x^4} - \frac{\delta_{jk} m_q}{4 \pi^2 x^2} - \frac{i t^a_{j k}
}{32 \pi^2 x^2}(\sigma^{\alpha \beta} x\!\!\!\slash
+ x\!\!\!\slash \sigma^{\alpha \beta}) - \frac{\delta_{jk}}{12} \langle\bar{q} q \rangle + \frac{i\delta_{j k}
x\!\!\!\slash}{48} m_q \langle \bar{q}q \rangle - \frac{\delta_{j k} x^2}{192} \langle g_s \bar{q} \sigma \cdot G q \rangle \nonumber \\ &+& \frac{i \delta_{jk} x^2 x\!\!\!\slash}{1152} m_q \langle g_s \bar{q} \sigma \cdot G q \rangle - \frac{t^a_{j k} \sigma_{\alpha \beta}}{192}
\langle g_s \bar{q} \sigma \cdot G q \rangle
+ \frac{i t^a_{jk}}{768} (\sigma_{\alpha \beta} x \!\!\!\slash + x\!\!\!\slash \sigma_{\alpha \beta}) m_q \langle
g_s \bar{q} \sigma \cdot G q \rangle \;,
\end{eqnarray}
\begin{eqnarray}
S^Q_{j k}(p) \! \! & = & \! \! \frac{i \delta_{j k}(p\!\!\!\slash + m_Q)}{p^2 - m_Q^2} - \frac{i}{4} \frac{t^a_{j k} }{(p^2 - m_Q^2)^2} [\sigma^{\alpha \beta}
(p\!\!\!\slash + m_Q)
+ (p\!\!\!\slash + m_Q) \sigma^{\alpha \beta}] \nonumber \\ &+& \frac{i\delta_{jk}m_Q  \langle g_s^2 G^2\rangle}{12(p^2 - m_Q^2)^3}\bigg[ 1 + \frac{m_Q (p\!\!\!\slash + m_Q)}{p^2 - m_Q^2} \bigg] \nonumber \\ &+& \frac{i \delta_{j k}}{48} \bigg\{ \frac{(p\!\!\!\slash +
m_Q) [p\!\!\!\slash (p^2 - 3 m_Q^2) + 2 m_Q (2 p^2 - m_Q^2)] }{(p^2 - m_Q^2)^6}
\times (p\!\!\!\slash + m_Q)\bigg\} \langle g_s^3 G^3 \rangle \; .
\end{eqnarray}
Here, the vacuum condensates are clearly displayed. For more explanation on above propagators, readers may refer to Refs.~\cite{Wang:2013vex, Albuquerque:2013ija}.

In the partonic representation, the dispersion relation may express the correlation function $\Pi(q^2)$ as:
\begin{eqnarray}
\Pi^{OPE} (q^2) &=& \int_{(2 m_b + m_q + m_{q^\prime})^2}^{\infty} d s
\frac{\rho^{OPE}(s)}{s - q^2} + \Pi^{sum}(q^2)\; ,
\label{OPE-hadron}
\end{eqnarray}
where $\rho^{OPE}(s) = \text{Im} [\Pi^{OPE}(s)] / \pi$ and
\begin{eqnarray}
\rho^{OPE}(s) & = & \rho^{pert}(s) + \rho^{\langle \bar{q} q
\rangle}(s) + \rho^{\langle \bar{q^\prime} q^\prime
\rangle}(s) + \rho^{\langle G^2 \rangle}(s) + \rho^{\langle \bar{q} G q \rangle}(s) \nonumber \\
&+& \rho^{\langle \bar{q^\prime} G q^\prime \rangle}(s)
+ \rho^{\langle \bar{q} q \rangle \langle \bar{q^\prime} q^\prime \rangle}(s) + \rho^{\langle G^3 \rangle}(s) + \rho^{\langle G^2 \rangle^2}(s)\; , \label{rho-OPE} \\
\Pi^{sum}(q^2) &=& \Pi^{\langle G^2 \rangle}(q^2) + \Pi^{\langle \bar{q} G q \rangle}(q^2) + \Pi^{\langle \bar{q^\prime} G q^\prime \rangle}(q^2) + \Pi^{\langle \bar{q} q \rangle \langle \bar{q^\prime} q^\prime \rangle}(q^2) \nonumber \\
&+& \Pi^{\langle G^3 \rangle}(q^2) + \Pi^{\langle \bar{q} q \rangle \langle \bar{q^\prime} G q^\prime \rangle}(q^2) + \Pi^{\langle \bar{q^\prime} q^\prime \rangle \langle \bar{q} G q \rangle}(q^2) + \Pi^{\langle G^2 \rangle^2}(q^2) \; . \label{rho-OPE-Pi}
\end{eqnarray}
Here, $\Pi^{sum }(q^2)$ is the sum of those contributions in the correlation function that have no imaginary part but have nontrivial magnitudes after the Borel transformatiom. By applying the Borel transformation to (\ref{OPE-hadron}), we have
\begin{eqnarray}
\Pi^{OPE}(M_B^2)\!\! = \!\!\int_{(2 m_Q + m_q + m_{q^\prime})^2}^{\infty} d s
\rho^{OPE}(s)e^{- s / M_B^2} + \Pi^{sum }(M_B^2)\ . \label{quark-gluon}
\end{eqnarray}
To consider the strange quark mass effect, those terms linear to the strange quark mass $m_s$ are kept in the calculation. The lengthy expressions of spectral densities in Eq.(\ref{quark-gluon}) are deferred to the Appendix.

In the hadronic representation, after isolating the ground state contribution from the tetraquark state, we obtain the correlation function $\Pi(q^2)$ in dispersion integral over the physical region, i.e.
\begin{eqnarray}
\Pi(q^2) & = & \frac{(\lambda^X)_2}{(M^X)^2 - q^2} + \frac{1}{\pi} \int_{s_0}^\infty d s \frac{\rho(s)}{s - q^2} \; , \label{hadron}
\end{eqnarray}
where the superscript $X$ means the lowest lying tetraquark state, $M^{X}$ denotes its mass, $\rho(s)$ is the spectral density that contains the contributions from higher excited states and the continuum states above the threshold $s_0$. The coupling constant $\lambda^X$ is defined through $\langle 0 | j_\mu | X \rangle = \lambda^X \epsilon_\mu$.

By performing the Borel transform on the hadronic side, the Eq.(\ref{hadron}), and matching it to Eq.(\ref{quark-gluon}), we can then obtain the mass of the tetraqark state
\begin{eqnarray}
M^X_{i-q^\prime}(s_0, M_B^2) = \sqrt{- \frac{L_1(s_0, M_B^2)}{L_0(s_0, M_B^2)}} \; , \label{mass-Eq}
\end{eqnarray}
where the subscripts $i$ runs from $A$ to $E$, $q^\prime = d$ or $s$, and the moments $L_1$ and $L_0$ are respectively defined as:
\begin{eqnarray}
L_0(s_0, M_B^2) & = & \int_{(2m_Q + m_q + m_{q^\prime})^2}^{s_0} d s \; \rho^{OPE}(s) e^{-
s / M_B^2} \nonumber \\
&+& \Pi^{sum}(M_B^2) \; , \label{L0}
\\ L_1(s_0, M_B^2) & = &
\frac{\partial}{\partial{\frac{1}{M_B^2}}}{L_0(s_0, M_B^2)} \; .
\end{eqnarray}

\section{Numerical Evaluation}

\subsection{Inputs}

In the QCD Sum Rule calculation, as in any practical theory, to yield meaningful physical results one needs to give certain inputs, such as the magnitudes of condensates and quark masses. Being ascertained in Refs.\cite{Shifman, Reinders:1984sr, P.Col, Narison:1989aq}, for numerical evaluation the inputs we take are:
$m_s(2 \, \text{GeV}) = (95 \pm 5) \; \text{MeV}$, $m_c (m_c) = \overline{m}_c= (1.275 \pm 0.025)
\; \text{GeV}$, $m_b (m_b) = \overline{m}_b= (4.18 \pm 0.03)
\; \text{GeV}$, $\langle \bar{q} q \rangle = - (0.24 \pm 0.01)^3
\; \text{GeV}^3$, $\langle \bar{s} s \rangle = (0.8 \pm 0.1)
\langle \bar{q} q \rangle$, $\langle g_s^2 G^2 \rangle = 0.88
\; \text{GeV}^4$, $\langle \bar{s} g_s \sigma \cdot G s
\rangle = m_0^2 \langle \bar{s} s \rangle$, $\langle g_s^3 G^3
\rangle = 0.045 \; \text{GeV}^6$, and $m_0^2 = 0.8 \; \text{GeV}^2$.
Here, ${\overline m}_c$ and ${\overline m}_b$ represent heavy-quark running masses in $\overline{MS}$ scheme. For light quarks $u$ and $d$, the chiral limit masses $m_u = m_d =0$ are adopted.

Moreover, there exist two additional parameters $M_B^2$ and $s_0$ introduced in establishing the sum rule, which will be fixed in light of the so-called standard procedures abiding by two criteria \cite{Shifman, Reinders:1984sr, P.Col}. The first one asks for the convergence of the OPE. That is, one needs to compare individual contributions with the total magnitude on the OPE side, and choose a reliable region for $M_B^2$ to retain the convergence. In practice, the degree of convergence may be expressed in fractions of condensates over the total as per
\begin{eqnarray}
  R_{i - q^\prime}^{OPE} = \frac{L_0^{dim}(s_0, M_B^2)}{L_0(s_0, M_B^2)}\, ,
\end{eqnarray}
where the superscript $dim$ means the dimension of relevant condensate in the OPE of Eq.(\ref{L0}), the subscript $i$ runs from $A$ to $E$, and $q^\prime = d$ or $s$.

The second criterion requires that the portion of lowest lying pole contribution (PC), the ground state contribution, in the total, pole plus continuum, should be over 50\%~\cite{P.Col, Matheus:2006xi}, which can be formulated as
\begin{eqnarray}
  R_{i - q^\prime}^{PC} = \frac{L_0(s_0, M_B^2)}{L_0(\infty, M_B^2)} \; . \label{RatioPC}
\end{eqnarray}
Here the subscripts $i$ runs from $A$ to $E$, and $q^\prime = d$ or $s$. Under this prerequisite, the contributions of higher excited and continuum states will be properly suppressed.

In order to find a proper value for continuum threshold $s_0$, we perform a similar analysis as in Refs.~\cite{Finazzo:2011he, Qiao:2013dda}. Notice that the $s_0$ relates to the mass of the ground state by $\sqrt{s_0} \sim (M_X + \delta) \, \text{GeV}$, in which $\delta$ lies in the scope of $0.4 \sim 0.8$ GeV, various $\sqrt{s_0}$ satisfying this constraint should be taken into account in the numerical analysis. Among these values, one needs to pick up the one which yields an optimal window for Borel parameter $M_B^2$. That is to say, in the optimal window, the tetraquark mass $M_X$ is somehow independent of the Borel parameter $M_B^2$. In this way, the value of $\sqrt{s_0}$ corresponding to the optimal mass curve will be taken as its central value. In practice, we may vary $\sqrt{s_0}$ by $0.2$ GeV in numerical calculation \cite{Qiao:2013raa}, which set the upper and lower bounds and hence the uncertainties of $\sqrt{s_0}$.

\subsection{The color-octet-octet configuration}

For case A, we illustrate the OPE convergence in Figs.(\ref{fig1}-a) and (\ref{fig1}-b) for $q^\prime = d$ and $s$, respectively. According to the first criterion, we find the lower limit of $M_B^2$, i.e., $M_B^2 \gtrsim 9.3 \; \text{GeV}^{2}$ ($M_B^2 \gtrsim 9.4 \; \text{GeV}^{2}$ ) with $\sqrt{s_0} = 12.0 \, \text{GeV}$ ($\sqrt{s_0} = 12.1 \, \text{GeV}$). The curve of the pole contribution $R_{A-d}^{PC}$ ($R_{A-s}^{PC}$) is drawn in Fig.(\ref{fig1}-c)(Fig.(\ref{fig1}-d)), which gives the upper bound for $M_B^2$, i.e., $M_B^2 \lesssim 12.6 \; \text{GeV}^{2}$ ($M_B^2 \lesssim 13.1 \; \text{GeV}^{2}$) with $\sqrt{s_0} = 12.0 \, \text{GeV}$ ($\sqrt{s_0} = 12.1 \, \text{GeV}$). It should be noted that the constraints on $M_B^2$ also depend on the threshold parameter $\sqrt{s_0}$. To determine the $\sqrt{s_0}$, we need to apply the criterion two \cite{Matheus:2006xi}. The masses $M^X_{A-d}$ and $M^X_{A-s}$ as functions of the Borel parameter $M_B^2$ for different $\sqrt{s_0}$ are drawn in Figs.(\ref{fig2}-a,\ref{fig2}-b).

For case B, the OPE convergences are shown in Figs.(\ref{fig3}-a) and (\ref{fig3}-b) for $q^\prime = d$ and $s$, respectively. According to the first criterion, we may find the lower limit of $M_B^2$, i.e., $M_B^2 \gtrsim 9.3 \; \text{GeV}^{2}$ ($M_B^2 \gtrsim 9.8 \; \text{GeV}^{2}$ ) with $\sqrt{s_0} = 12.0 \, \text{GeV}$ ($\sqrt{s_0} = 12.1 \, \text{GeV}$). The curve of the pole contribution $R_{B-d}^{PC}$ ($R_{B-s}^{PC}$) is drawn in Fig.(\ref{fig3}-c)(Fig.(\ref{fig3}-d)), which give out the upper bound for $M_B^2$, i.e., $M_B^2 \lesssim 12.6 \; \text{GeV}^{2}$ ($M_B^2 \lesssim 13.0 \; \text{GeV}^{2}$) with $\sqrt{s_0} = 12.0 \, \text{GeV}$ ($\sqrt{s_0} = 12.1 \, \text{GeV}$). The masses $M^X_{B-d}$ and $M^X_{B-s}$ as functions of the Borel parameter $M_B^2$ for different values $\sqrt{s_0}$ are depicted in Figs.(\ref{fig4}-a,\ref{fig4}-b).

\begin{widetext}

\begin{figure}
\begin{center}
\includegraphics[width=6.5cm]{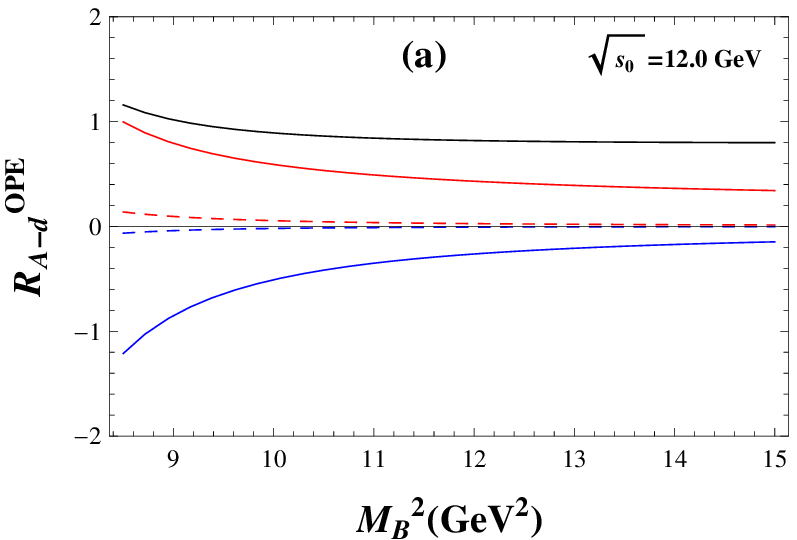}
\includegraphics[width=6.5cm]{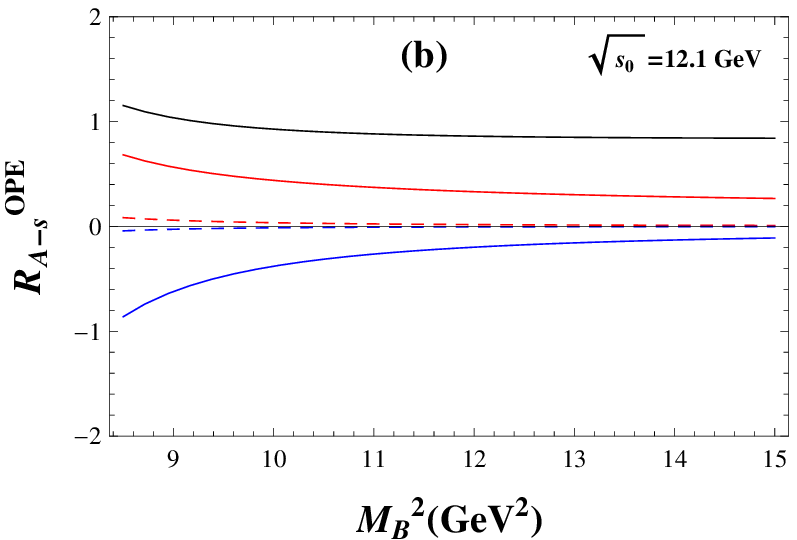}
\includegraphics[width=6.8cm]{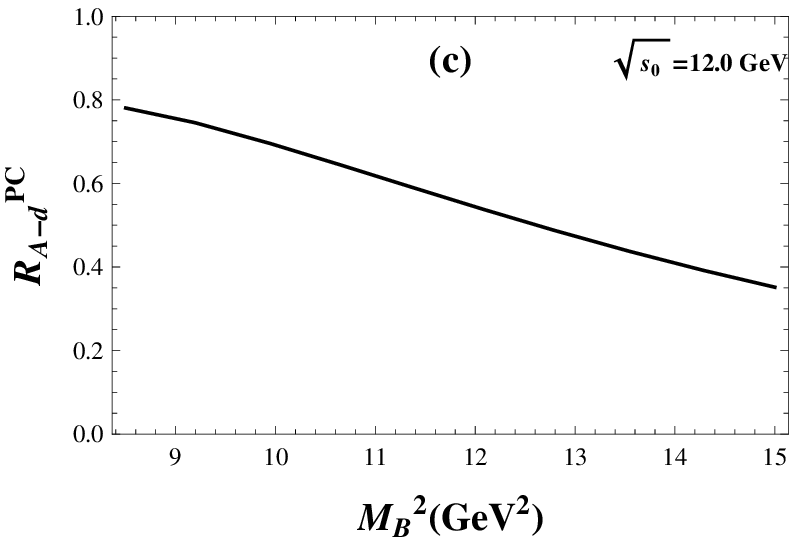}
\includegraphics[width=6.8cm]{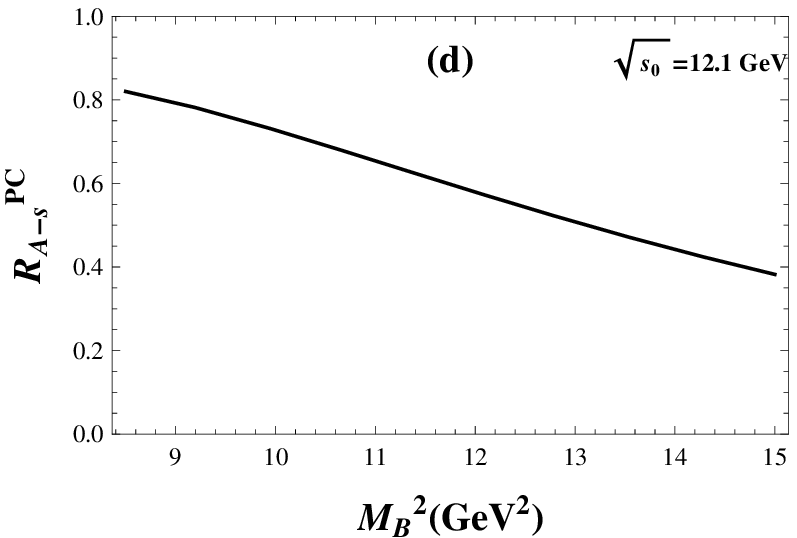}
\caption{The case A evaluation results. (\textbf{a}) The OPE convergence $R^{OPE}_{A-d}$ as function of the Borel parameter $M_B^2$ in the region $8.5 \leq M_B^2 \leq 15.0 \; \text{GeV}^{2}$ with $q^\prime = d$ and $\sqrt{s_0} = 12.0 \; \text{GeV}$, where the solid, red, blue, red-dashed, and blue-dashed lines represent the ratios of the contributions from perturbative, $\langle \bar{q} q \rangle$, $\langle \bar{q} G q \rangle$, $\langle \bar{q} q \rangle^2$, and $\langle \bar{q} q\rangle \langle \bar{q} G q\rangle$ terms to moment $L_0 (s_0, M_B^2)$, respectively. The $\langle GG\rangle$ and $\langle GG\rangle^2$ terms contributions are not displayed, since their magnitudes are tiny. (\textbf{b}) The OPE convergence $R^{OPE}_{A-s}$ as function of the Borel parameter $M_B^2$ in the region $8.5 \leq M_B^2 \leq 15.0 \; \text{GeV}^{2}$ with $q^\prime = s$ and $\sqrt{s_0} = 12.1 \; \text{GeV}$. (\textbf{c}) The pole contribution $R^{PC}_{A-d}$ as function of the Borel parameter $M_B^2$ with $q^\prime = d$ and $\sqrt{s_0} = 12.0 \; \text{GeV}$. (\textbf{d}) The pole contribution $R^{PC}_{A-s}$ as function of the Borel parameter $M_B^2$ with $q^\prime = s$ and $\sqrt{s_0} = 12.1 \; \text{GeV}$.} \label{fig1}
\end{center}
\end{figure}
\begin{figure}
\begin{center}
\includegraphics[width=6.8cm]{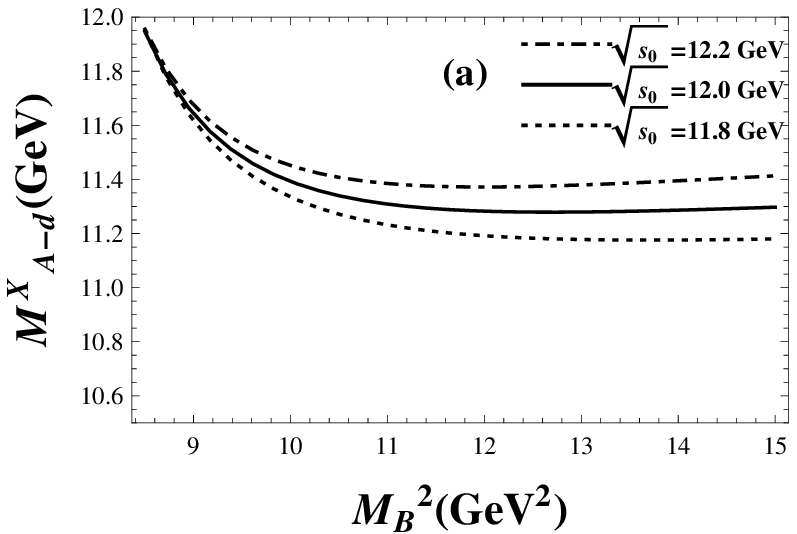}
\includegraphics[width=6.8cm]{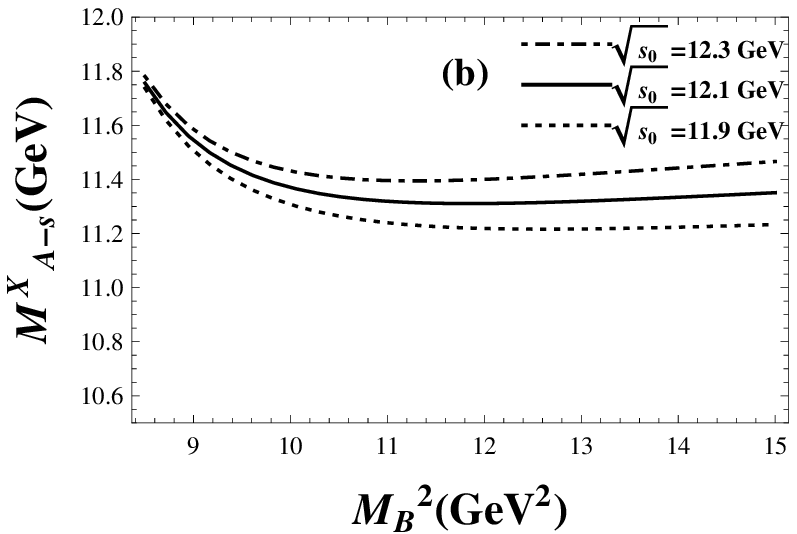}
\caption{(\textbf{a}) The tetraquark mass $M^X_{A-d}$ as function of the Borel parameter $M_B^2$ of case $A$ with $q^\prime = d$ and different values of $\sqrt{s_0}$. (\textbf{b}) The tetraquark mass $M^X_{A-s}$ as function of the Borel parameter $M_B^2$ of case $A$ with $q^\prime = s$ and different values of $\sqrt{s_0}$.} \label{fig2}
\end{center}
\end{figure}

\begin{figure}
\begin{center}
\includegraphics[width=6.8cm]{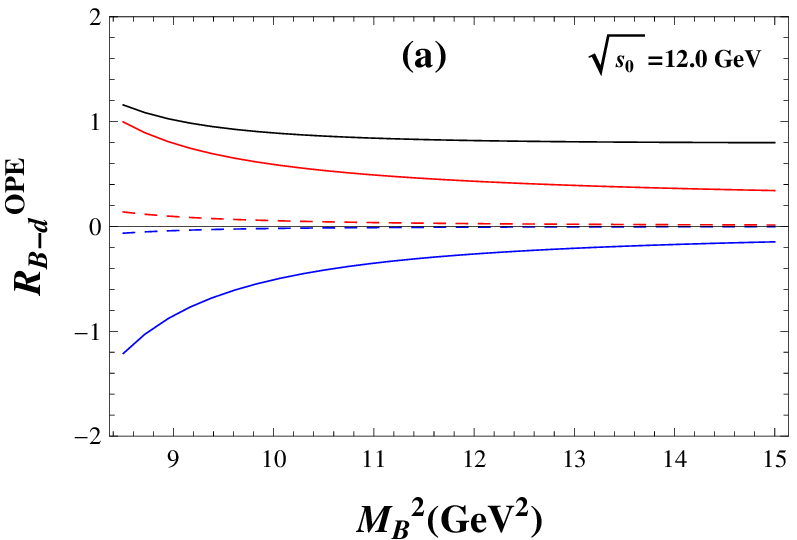}
\includegraphics[width=6.8cm]{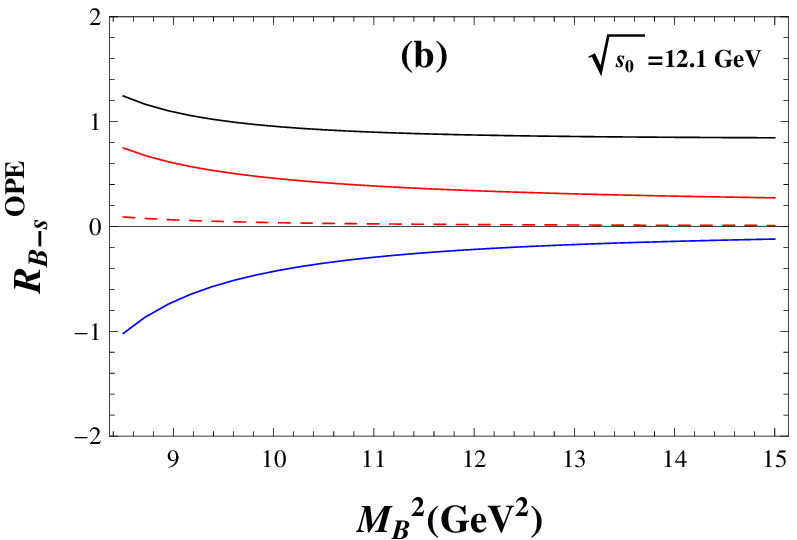}
\includegraphics[width=6.8cm]{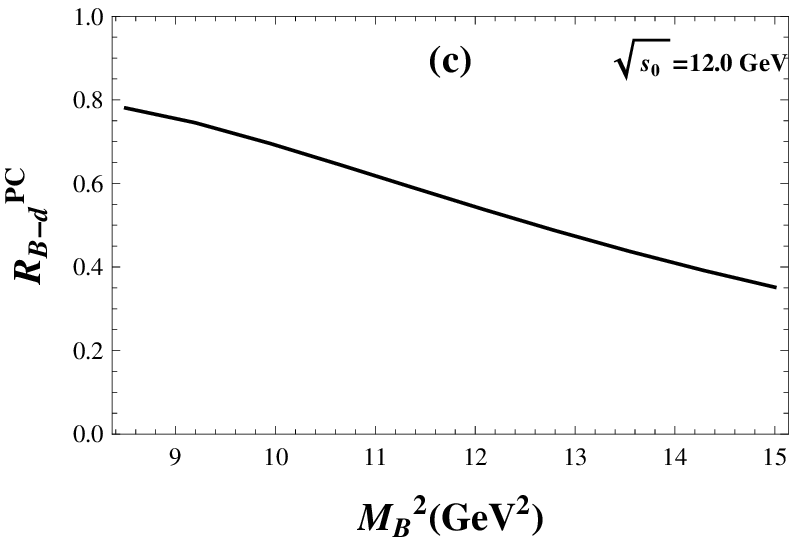}
\includegraphics[width=6.8cm]{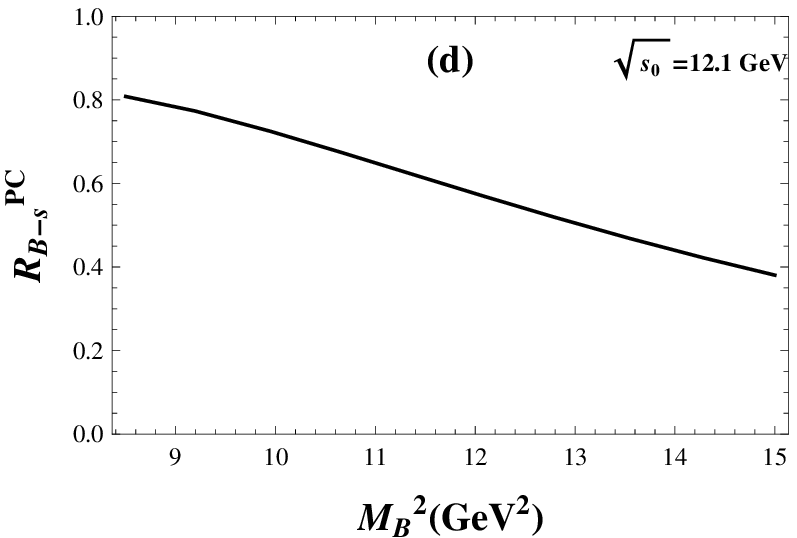}
\caption{Similar captions as in Fig. \ref{fig1} for case $B$. } \label{fig3}
\end{center}
\end{figure}
\begin{figure}
\begin{center}
\includegraphics[width=6.8cm]{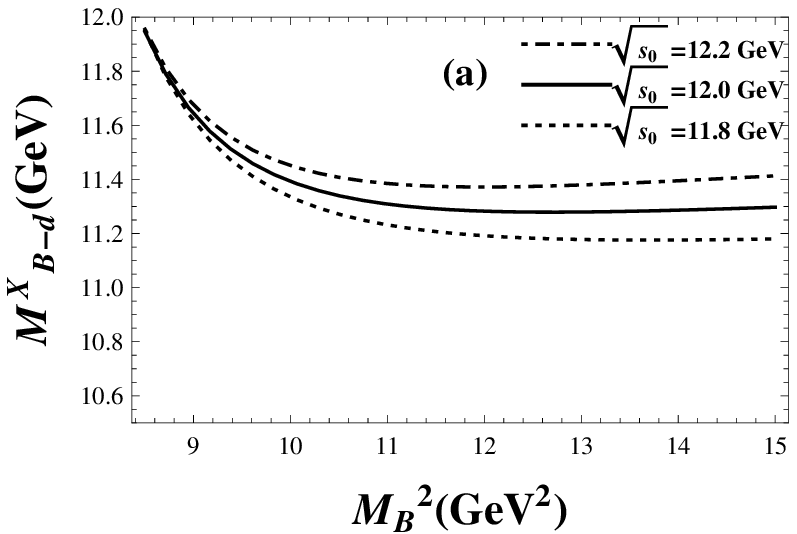}
\includegraphics[width=6.8cm]{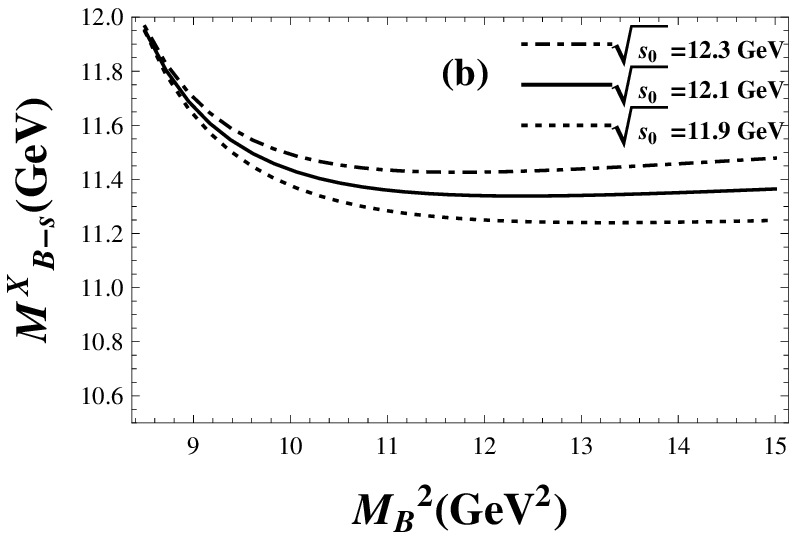}
\caption{Similar captions as in Fig. \ref{fig2} for case $B$.} \label{fig4}
\end{center}
\end{figure}

\begin{figure}
\begin{center}
\includegraphics[width=6.cm]{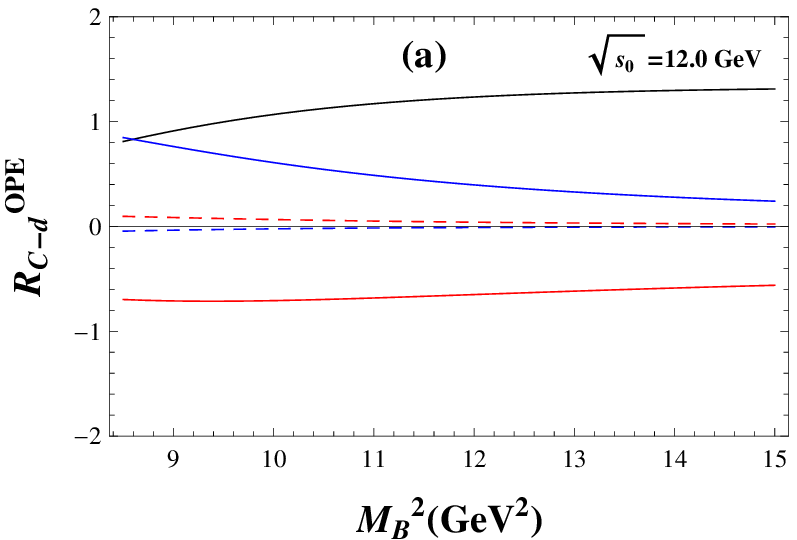}
\includegraphics[width=6.8cm]{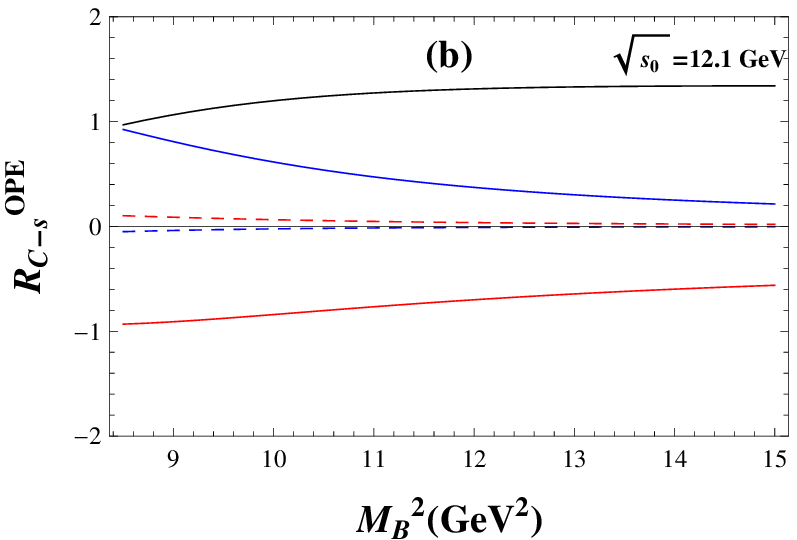}
\includegraphics[width=6.8cm]{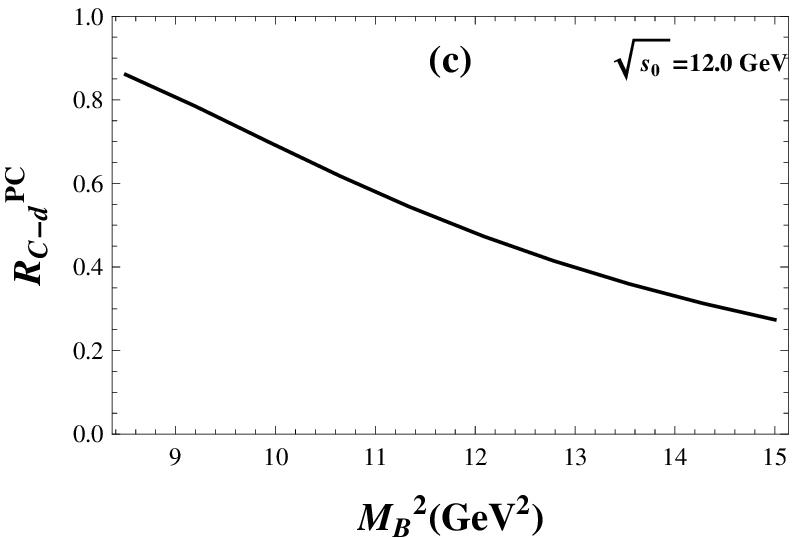}
\includegraphics[width=6.8cm]{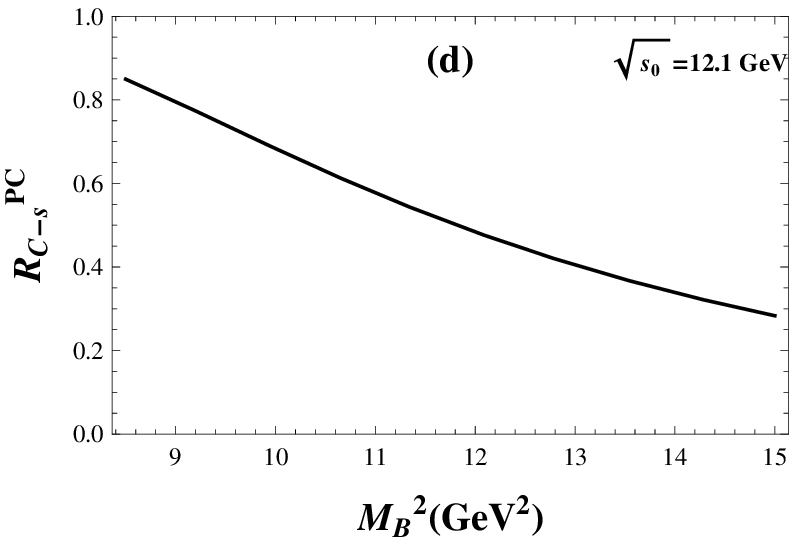}
\caption{Similar captions as in Fig. \ref{fig1} for case $C$.} \label{fig5}
\end{center}
\end{figure}
\begin{figure}
\begin{center}
\includegraphics[width=6.8cm]{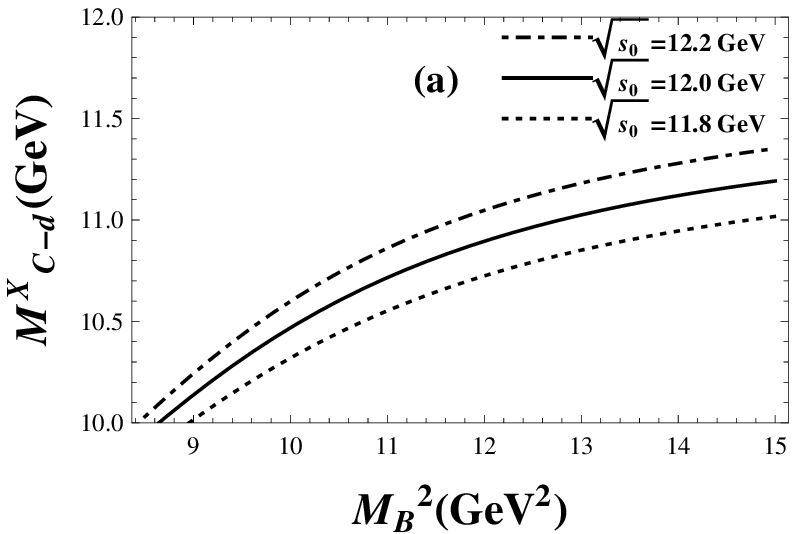}
\includegraphics[width=6.8cm]{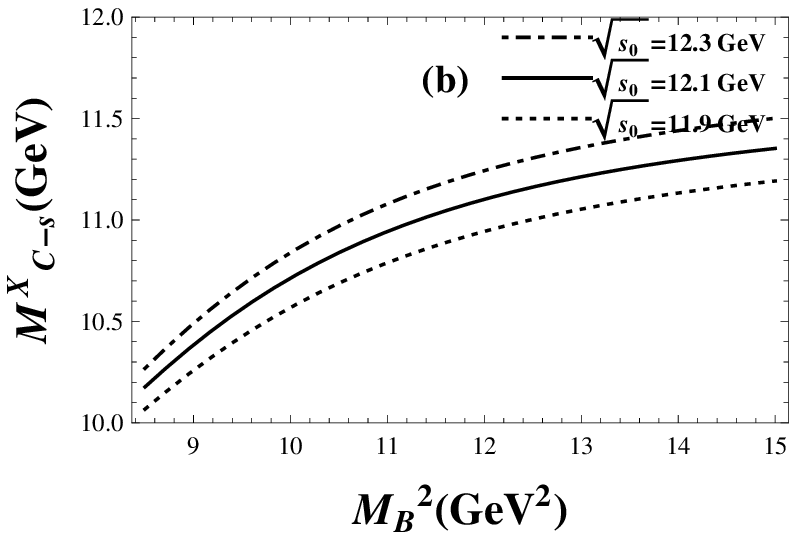}
\caption{Similar captions as in Fig. \ref{fig2} for case $C$.} \label{fig6}
\end{center}
\end{figure}

\begin{figure}
\begin{center}
\includegraphics[width=6.8cm]{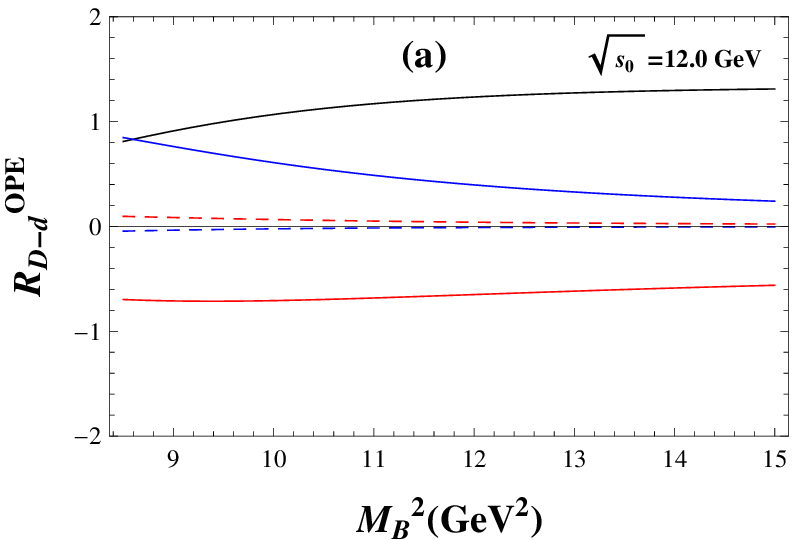}
\includegraphics[width=6.8cm]{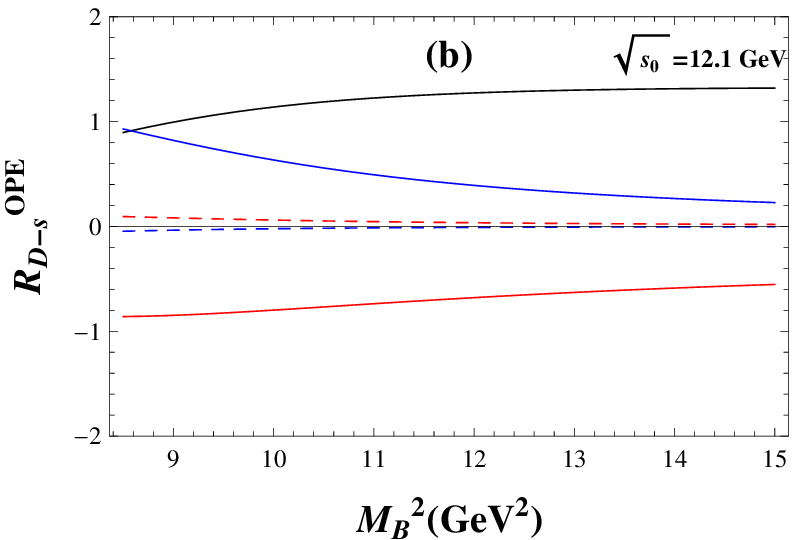}
\includegraphics[width=6.8cm]{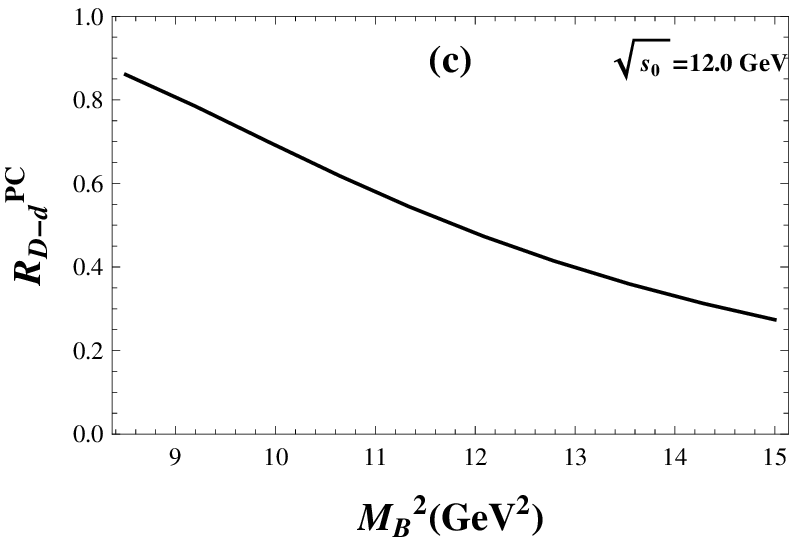}
\includegraphics[width=6.8cm]{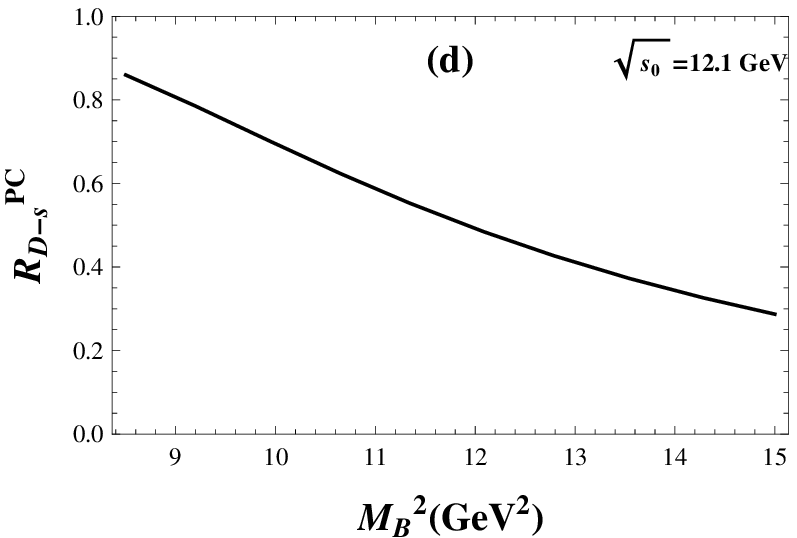}
\caption{Similar captions as in Fig. \ref{fig1} for case $D$.} \label{fig7}
\end{center}
\end{figure}
\begin{figure}
\begin{center}
\includegraphics[width=6.8cm]{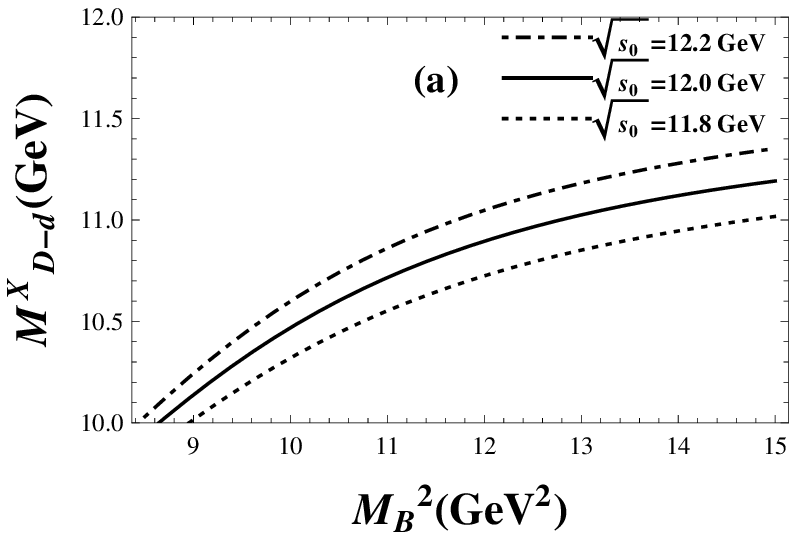}
\includegraphics[width=6.8cm]{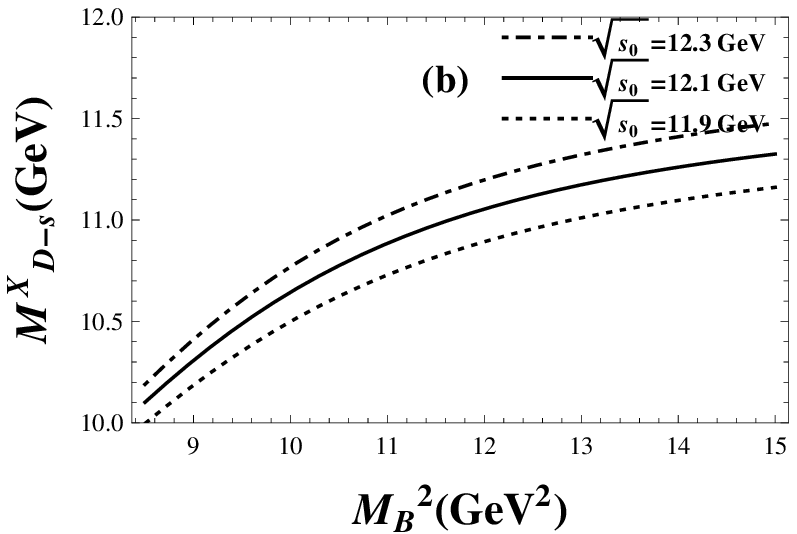}
\caption{Similar captions as in Fig. \ref{fig2} for case $D$.} \label{fig8}
\end{center}
\end{figure}

\begin{figure}
\begin{center}
\includegraphics[width=6.8cm]{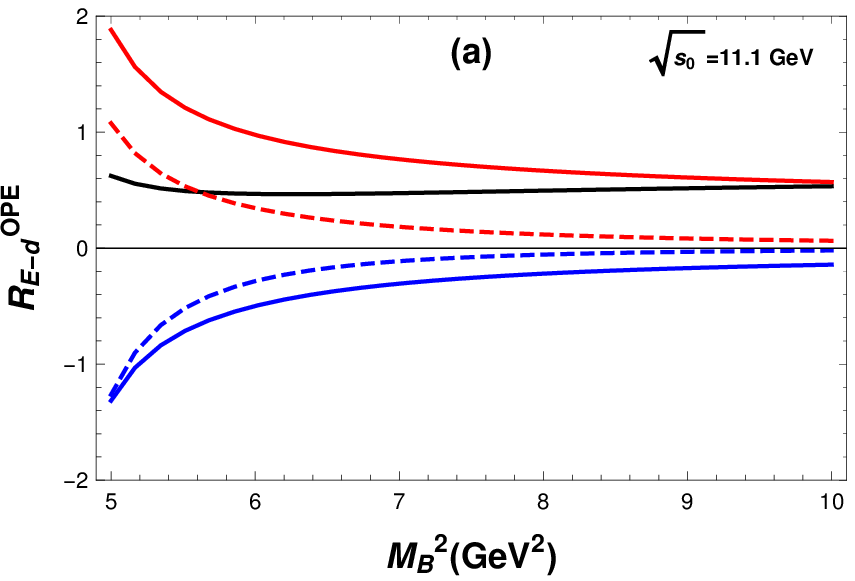}
\includegraphics[width=6.8cm]{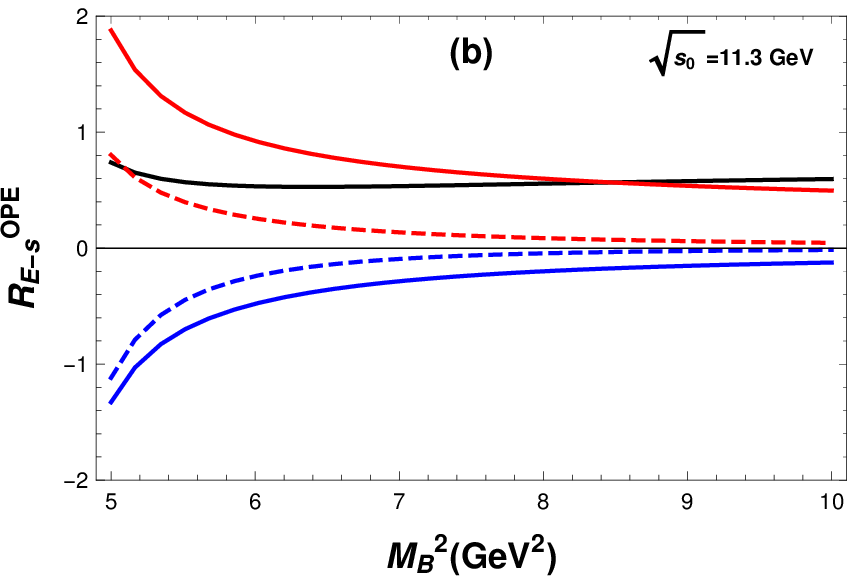}
\includegraphics[width=6.8cm]{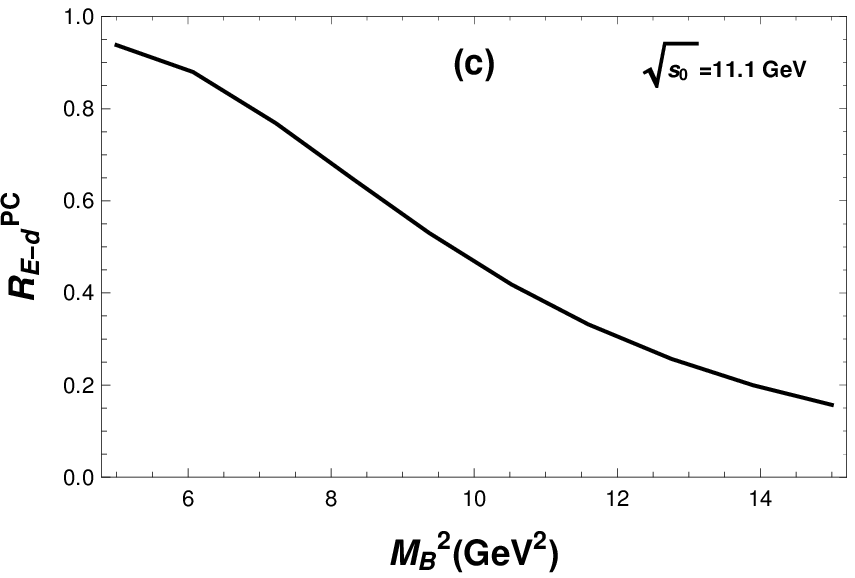}
\includegraphics[width=6.8cm]{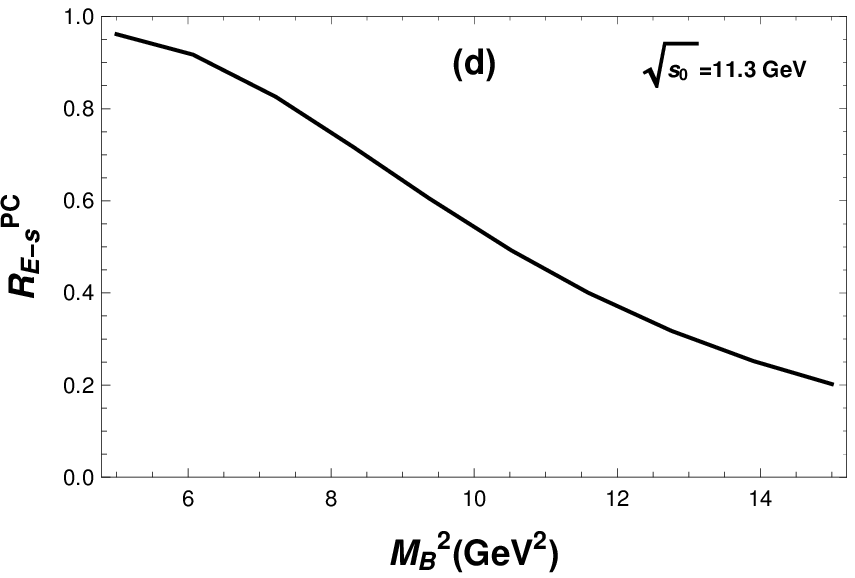}
\caption{The same caption as in Fig. \ref{fig1}, but for case $E$.} \label{fig9}
\end{center}
\end{figure}

\begin{figure}
\begin{center}
\includegraphics[width=6.8cm]{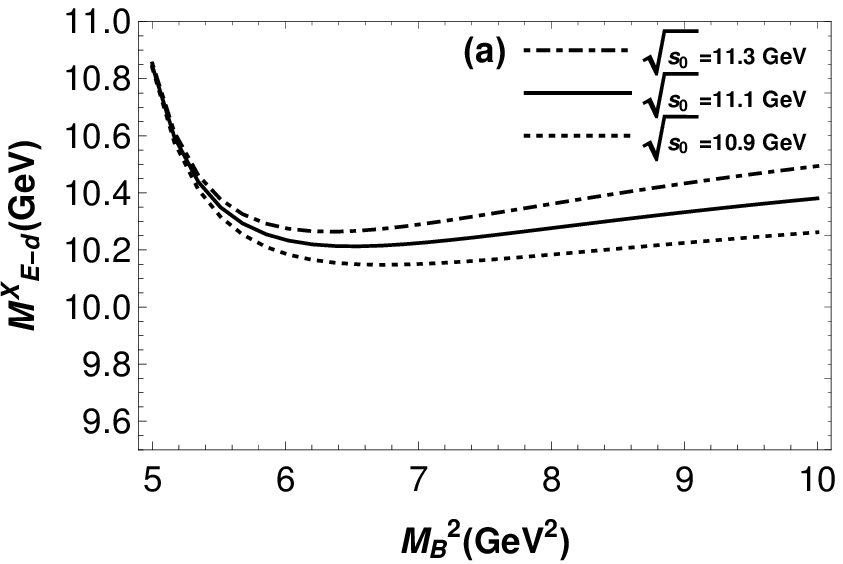}
\includegraphics[width=6.8cm]{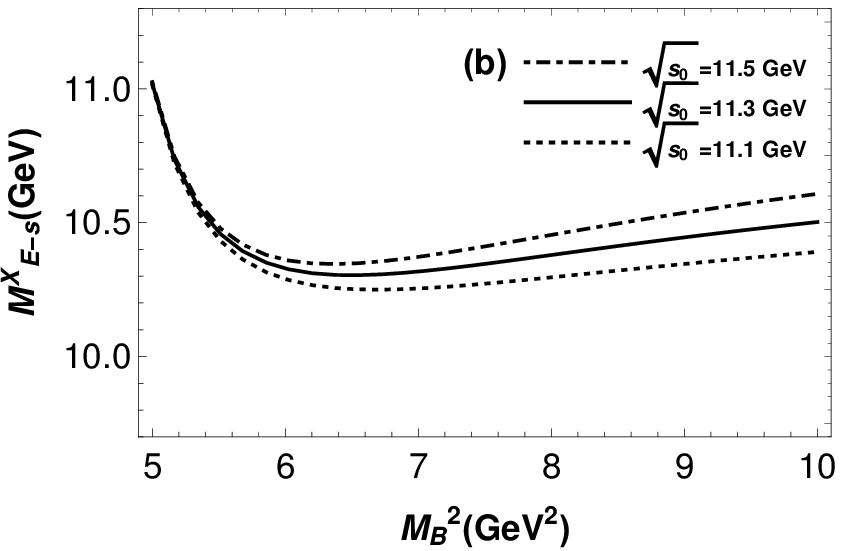}
\caption{The same caption as in Fig. \ref{fig2}, but for case $E$.} \label{fig10}
\end{center}
\end{figure}

\end{widetext}

Note, for cases $C$ and $D$, however, we can not find reasonable parameters $M_B^2$, which implies that those two currents have no strong coupling with the tetraquark states.

After the above evaluation, we can then determine the masses of the tetraquark states with currents $A$ and $B$, that is
\begin{eqnarray}
M_{A-d}^X &=& (11.28 \pm 0.15) \, \text{GeV} \; , \\
M_{A-s}^X &=& (11.31 \pm 0.16) \, \text{GeV} \; , \\
M_{B-d}^X &=& (11.28 \pm 0.15) \, \text{GeV} \; , \\
M_{B-s}^X &=& (11.34 \pm 0.16) \, \text{GeV} \; ,\label{eq-mass-1}
\end{eqnarray}
where the central values correspond to the results with the optimal stability of $M_B^2$, and the errors stem from the uncertainties of the condensates, the quark mass, the threshold parameter $\sqrt{s_0}$, and the Borel parameter $M_B^2$.

Replace $m_b$ with $m_c$ in the correlation functions and perform the same procedure on justifying the Borel parameter $M_B^2$ and threshold $\sqrt{s_0}$, we may obtain results for doubly charmed systems. That is $(4.72 \pm 0.13)$ for the $u d \bar{c} \bar{c}$ system, and $(4.76 \pm 0.14)$ and $(4.78 \pm 0.14)$ GeV for the $u s \bar{c} \bar{c}$ system.

\subsection{Molecular configuration}

The OPE convergence in molecular configuration is shown in Figs.(\ref{fig9}-a, \ref{fig9}-b) for $q^\prime = d$ and $s$ respectively. Considering the first criterion in setting up the sum rule, we find the lower limit $M_B^2$ goes as $M_B^2 \gtrsim 6.0 \; \text{GeV}^{2}$ ($M_B^2 \gtrsim 6.0 \; \text{GeV}^{2}$ ) with $\sqrt{s_0} = 11.1 \, \text{GeV}$ ($\sqrt{s_0} = 11.3 \, \text{GeV}$) for $q^\prime = d$ ($q^\prime = s$). The portion of pole contribution $R_{E-d}^{PC}$ ($R_{E-s}^{PC}$) is shown in Fig.(\ref{fig9}-c)(Fig.(\ref{fig9}-d)), which determines the upper bound for $M_B^2$ as $M_B^2 \lesssim 9.5 \; \text{GeV}^{2}$ ($M_B^2 \lesssim 10.5 \; \text{GeV}^{2}$) with $\sqrt{s_0} = 11.1 \, \text{GeV}$ ($\sqrt{s_0} = 11.3 \, \text{GeV}$) for $q^\prime = d$ ($q^\prime = s$). Here, the threshold $\sqrt{s_0}$ is chosen to be the optimal value that guarantees the stability of the calculated mass on $M_B^2$. The $M^X_{E-d}$ and $M^X_{E-s}$ dependence on Borel parameter $M_B^2$ for different $\sqrt{s_0}$ are shown in Figs.(\ref{fig10}-a,\ref{fig10}-b). In the end of the evaluation, we find the most acceptable masses of the tetraquark in molecular configuration are
\begin{eqnarray}
  M^X_{E-d} &=& (10.36 \pm 0.15) \, \text{GeV} \, ,\nonumber \\
  M^X_{E-s} &=& (10.48 \pm 0.15) \,  \text{GeV} \, . \label{mass-E}
\end{eqnarray}

For the charm sector, taking the same procedure as in $b$ sector with $m_b$ being replaced by $m_c$ in the spectral density, the masses of $ud\bar{c}\bar{c}$ and $us\bar{c}\bar{c}$ in molecular configuration are then obtained as $(4.04 \pm 0.13)$, and $(4.18 \pm 0.13)$, respectively. Note that Ref.\cite{Francis:2016hui} finds that the charm quark mass is not so large as to take the $m_Q \to \infty$ limit and the stability of the doubly charmed tetraquark structure is questionable.

\begin{table}
\begin{center}
\begin{tabular}{|c|c|c|c|c|c}\hline\hline
& $M_B^2 (\rm{GeV}^2)$ & $\sqrt{s_0} (\rm{GeV})$ & pole  & $M^X$ (\rm{GeV})   \\ \hline
 case $A\!\!-\!\!d$ & $9.3-12.6$   & $12.0\pm 0.2$  & $(69-50)\%$   & $11.28 \pm 0.15 $ \\ \hline
 case $A\!\!-\!\!s$ & $9.4-13.1$   & $12.1\pm 0.2$   & $(76-50)\%$  & $11.31 \pm 0.16 $       \\ \hline
 case $B\!\!-\!\!d$ & $9.3-12.6$   & $12.0\pm 0.2$  & $(69-50)\%$   & $11.28 \pm 0.15 $ \\ \hline
 case $B\!\!-\!\!s$ & $9.8-13.0$   & $12.1\pm 0.2$   & $(75-50)\%$  & $11.34 \pm 0.16 $       \\ \hline
 case $E\!\!-\!\!d$ & $5.5-9.6$   & $11.1\pm 0.2$   & $(92-50)\%$  & $10.36 \pm 0.15 $       \\ \hline
 case $E\!\!-\!\!s$ & $5.6-10.4$   & $11.3\pm 0.2$   & $(94-50)\%$  & $10.48 \pm 0.16 $       \\ \hline
 \hline
\end{tabular}
\end{center}
\caption{The Borel parameters, continuum thresholds, pole contributions, and predicted masses of the tetraquark states with currents $A$, $B$ and $E$.}
\label{tab1}
\end{table}

For the convenience of reference, a collection of Borel parameters, continuum thresholds, pole contributions and predicted masses for cases $A$, $B$ and $E$ are tabulated in Table~\ref{tab1}.

\subsection{The decay widths of the color-octet-octet tetraquark states}

Eq.(\ref{eq-mass-1}) tells that the masses of the double-bottom color-octet-octet tetraquarks are higher than the open bottom thresholds. Hence, they could decay to two bottom mesons through strong interaction, which implies their decay widths might be large. Following we calculate the decay widths by means of the QCD sum rules.

The calculation of the decay vertex starts from the three-point function:
\begin{eqnarray}
  \Pi_{\mu \nu} (q, q_1, q_2) = \int d^4x d^4y e^{iq_1 \cdot x} e^{i q_2 \cdot y} \Pi_{\mu \nu}(x,y)\ , \label{three-point}
\end{eqnarray}
where $\Pi_{\mu \nu}(x,y )= \langle 0|T[j_\mu^{B^*}(x) j_5^{B^0}(y) j_\nu^{X \dagger}(0)]|0\rangle$ and $q = q_1 + q_2$. The interpolating currents of $B^*$ and $B^0$ have the following forms:
\begin{eqnarray}
  j_\mu^{B^*} &=& \bar{b}_a \gamma_\mu u_a \, , \\
  j_5^{B^0} &=& i \bar{b}_b \gamma_5 d_b \ .
\end{eqnarray}

On the phenomenological side of the QCD sum rules, we insert the intermediate states into Eq.(\ref{three-point}), and get
\begin{eqnarray}
      \Pi_{\mu \nu}^{phen} &=& \frac{ -\lambda_X m_{B^*} f_{B^*} f_{B_0} m_{B^0}^2 g_{X B^\ast B^0}(q^2)}{m_b (q^2 - M_X^{2})(q_1^2 - m_{B^*}^2) (q_2^2 - m_{B^0}^2)} \nonumber \\
      &\times& \bigg(- g_{\mu \alpha} + \frac{q_{1\mu} q_{1\alpha}}{m_{B^*}^2}\bigg)\bigg(-g_{\nu}^{\alpha} + \frac{q_{\nu} q^{\alpha}}{M_X^{2}}\bigg) ,
\end{eqnarray}
where $M_X$ is the mass of the tetraquark state, $\lambda_X$ is its decay constant, and $g_{X B^\ast B^0}(q^2)$ is the form factor. They are defined as follows:
\begin{eqnarray}
      \langle B^\ast B^0 | X \rangle &=& g_{X B^\ast B^0}(q^2) \varepsilon^\ast_\alpha(q_1) \varepsilon^\alpha(q) \; ,\\
      \langle 0 | j_\mu^{B^\ast} | B^\ast \rangle &=& m_{B^\ast} f_{B^\ast} \varepsilon_\mu(q_1) \; ,\\
      \langle 0 | j_\mu^{B^0} | B^0 \rangle &=& \frac{f_{B^0} m_{B^0}^2}{m_b} \; ,\\
      \langle X | j^{A, 1^+}_\nu(x)^\dagger | 0 \rangle &=& \lambda \varepsilon^\ast_\nu(q) \;.
\end{eqnarray}

\begin{figure}
\begin{center}
\includegraphics[width=6.8cm]{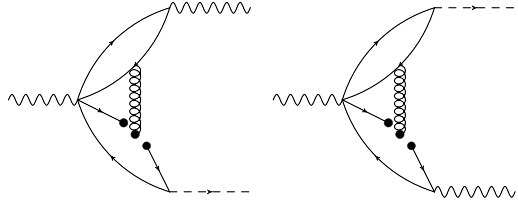}
\caption{The Feynman diagrams for calculating the decay width, where the wavy line represents a meson with $J = 1$ and the dashed line denotes a pseudoscalar meson.} \label{feyn-decay}
\end{center}
\end{figure}

On the OPE side, up to dimension five, the only nonzero term is the $\langle g_s \bar{q} \sigma \cdot G q \rangle$. As discussed in Ref.\cite{Dias:2013}, the $q_1^\mu q_1^\nu$ structure would give the dominant contribution, and hence we merely calculate those diagrams shown in Fig.(\ref{feyn-decay}) and then extract the coefficient of the $q_1^\mu q_1^\nu$ term. In the end, we have
\begin{eqnarray}
      \Pi^{OPE} &=& \frac{m_b \langle g_s \bar{q} \sigma \cdot G q \rangle}{96 \pi^2} \bigg[ \frac{1}{m_b^2 - q_2^2} \int^1_0 d \alpha \frac{1}{m_b^2 - ( 1 - \alpha ) q_1^2} \nonumber \\
      &+& \frac{1}{m_b^2 - q_1^2} \int^1_0 d \alpha \frac{1}{m_b^2 - (1 - \alpha) q_2^2} \bigg] \; , \label{OPE}
\end{eqnarray}

Performing Borel transforms on both $q^2$ and $q_1^2 $, we obtain
\begin{eqnarray}
      \frac{1}{Q^2 + m_{B^0}^2} &\bigg[& {\cal A} \left( e^{- m_{B^\ast}^2/M^2} - e^{-m_X^2/M^2} \right) +  B e^{-s_0/M^2} \bigg] \nonumber \\
      &=& \frac{m_b \langle g_s \bar{q} \sigma \cdot G q \rangle}{96 \pi^2}  \times \bigg[ \frac{1}{m_b^2 + Q^2} \int^1_0 d \alpha \frac{1}{1 - \alpha}  e^{\frac{-m_b^2}{(1 - \alpha) M^2}} \nonumber \\
      &+& e^{-m_b^2/M^2} \int^1_0 d \alpha \frac{1}{m_b^2 + (1 - \alpha) Q^2} \bigg] \; , \label{borel}
\end{eqnarray}
where $Q^2 = -q_2^2$ and $s_0$ is the continuum threshold of the tetraquark state, and $\mathcal{A}$ is defined as
\begin{eqnarray}
     {\cal A} = \frac{g_{X B^\ast B^0} \lambda_X f_{B^\ast} f_{B^0} m_{B^0}^2}{m_b m_{B^\ast} (m_X^2 - m_{B^\ast}^2) }\ .
\end{eqnarray}

In numerical evaluation, we adopt the mass and decay constant values of the bottom mesons given in Refs.\cite{Baker:2014, pdg}, i.e.,
\begin{eqnarray}
\begin{aligned}
& m_{B^\ast} = 5.33 \, \text{GeV} \; ,  & & f_{B^\ast} = 0.180 \, \text{GeV} \; ; \\
& m_{B^0} = 5.28 \, \text{GeV} \;, & & f_{B^0} = 0.180 \, \text{GeV} \; ; \\
& m_{B^\ast_s} = 5.42 \, \text{GeV} \; , & & f_{B^\ast_s} = 0.259 \, \text{GeV} \; ;  \\
& m_{B^0_s} = 5.37 \, \text{GeV} \; , & & f_{B^0_s} = 0.232 \, \text{GeV} \;.
\end{aligned}
\end{eqnarray}
Given the Borel window satisfying the two criteria mentioned above, we can get $g_{X B^\ast B^0}=g_{X B^\ast B^0}(-m_{B^0}^2)=4.49 \pm 0.41 \text{GeV}$.

The decay width takes the form
\begin{eqnarray}
      \Gamma_{A-d} (X \to B^\ast B^0) &=& \frac{p^*(m_X,m_{B^\ast},m_{B^0})}{8 \pi m_X^2} \nonumber \\
      &\times& \frac{g_{X B^\ast B^0}^2}{3} \left( 3 + \frac{(p^*(m,m_{B^\ast},m_{B^0}))^2}{m_{B^\ast}^2}  \right) \; ,
\end{eqnarray}
where the subscript $A-d$ stands for case $A$ with $q^\prime = d$, and
\begin{eqnarray}
      p^*(a,b,c)=\frac{\sqrt{a^4 + b^4 + c^4 - 2 a^2 b^2 - 2 a^2 c^2 - 2 b^2 c^2}}{2a} \; .
\end{eqnarray}
With the preceding inputs, we then get
\begin{eqnarray}
&&\Gamma_{A-d} = (12.60 \pm 2.20) \, \text{MeV}  \; . \\
&&\Gamma_{A-s} = (5.78 \pm 2.00) \, \text{MeV}  \; , \\
&&\Gamma_{B-d} = (12.60 \pm 2.20) \, \text{MeV}  \; , \\
&&\Gamma_{B-s} = (4.76 \pm 2.00) \, \text{MeV}  \; ,
\end{eqnarray}
where the subscript $A-s$, $B-d$, and $B-s$ denote case $A$ with $q^\prime = s$, case $B$ with $q^\prime = d$, and $q^\prime =s$, respectively.

\subsection{The Tetraquark mass dependence on radiative corrections}

In order to estimate the radiative correction effect on the leading order calculation, we define the ratios of radiative correction over the corresponding leading order ones as $\mathcal{A} \alpha_s(M_B^2)$, $\mathcal{B} \alpha_s(M_B^2)$, $\mathcal{C} \alpha_s(M_B^2)$, $\mathcal{D} \alpha_s(M_B^2)$, and $\mathcal{E} \alpha_s(M_B^2)$ for the perturbative, two-quark condensate, mixed condensate and four-quark condensate, and dimension-8 condensate composed of the two-quark condensate and mixed condensates in Eqs.(\ref{rho-OPE}, \ref{rho-OPE-Pi}), respectively, as
\begin{eqnarray}
\rho^{\prime pert}& =& (1 + \mathcal{A} \alpha_s) \rho^{pert} , \\
\rho^{\prime \langle \bar{q} q \rangle} &=& (1 + \mathcal{B} \alpha_s) \rho^{\langle \bar{q} q \rangle} , \\
\rho^{\prime \langle \bar{q} G q \rangle} &=& (1 + \mathcal{C} \alpha_s) \rho^{\langle \bar{q} G q \rangle} ,\\
\rho^{\prime  \langle \bar{q} q \rangle^2} &=& (1 + \mathcal{D} \alpha_s) \rho^{\langle \bar{q} q \rangle^2} , \\
\rho^{\prime \langle \bar{q} q\rangle \langle \bar{q} G q\rangle} &=& (1 + \mathcal{E} \alpha_s) \rho^{\langle \bar{q} q\rangle \langle \bar{q} G q\rangle} .\
\end{eqnarray}
Here, numerically $\alpha_s(M_B^2)=0.227$. Suppose the higher order contributions are subdominant comparing to the leading-order ones, the coefficients $\mathcal{A} \alpha_s(M_B^2)$, $\mathcal{B} \alpha_s(M_B^2)$, $\mathcal{C} \alpha_s(M_B^2)$, $\mathcal{D} \alpha_s(M_B^2)$, and $\mathcal{E} \alpha_s(M_B^2)$  are legitimately restricted the range of $[-1,1]$, which means $\mathcal{A} $, $\mathcal{B} $, $\mathcal{C} $, $\mathcal{D}$, and $\mathcal{E}$  fall into $[-4.4, 4.4]$.

\begin{figure}
\begin{center}
\includegraphics[width=8.8cm]{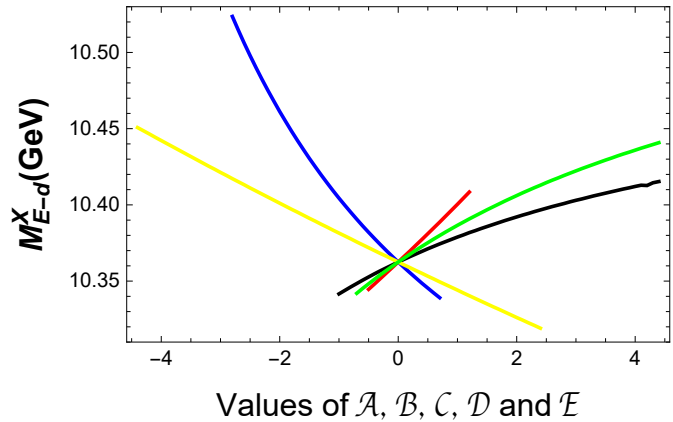}
\caption{The mass $M^X_{E-d}$ versus radiative corrections. Here the black, blue, red, yellow, and green lines represent the $M^X_{E-d}$ dependence on factors $\mathcal{A} $, $\mathcal{B} $, $\mathcal{C} $, $\mathcal{D}$, and $\mathcal{E}$. } \label{mass_dependent_on_alphas}
\end{center}
\end{figure}

Taking $M^X_{E-d}$ as an example, we adjust the parameter $\mathcal{A}$ to find the reasonable Borel window and get the valid regions in order to sustain the tetraquark existance. We find that to ensure a reasonable Borel window, while $\mathcal{A} \in [-1, 4.4]$, the uncertainty of $M^X_{E-d}$ due to the radiative correction of the perturbative term is confined to the range of -0.021 GeV to 0.053 GeV. We take as well the same procedures to coefficients $\mathcal{B}$, $\mathcal{C}$, $\mathcal{D}$, and $\mathcal{E}$ to find legitimate Borel windows, which are $\mathcal{B} \in [-2.8, 0.8]$, $\mathcal{C} \in [-0.5, 1.2]$, $\mathcal{D} \in [-4.4, 2.4]$, and $\mathcal{E} \in [-0.7, 4.4]$. The corresponding corrections to the mass $M^X_{E-d}$ therefore are $[-0.023, 0.161]$ GeV, $[-0.018, 0.050]$ GeV, $[-0.043, -0.088]$ GeV, and $[-0.021, 0.078]$, respectively, which indicates that the uncertainties induced by radiative corrections are quite small. The dependences of $M^X_{E-d}$ on radiative corrections are depicted in Fig.(\ref{mass_dependent_on_alphas}).

\section{Conclusions}

The doubly heavy tetraquarks are very interesting and important topics in the study of new hadronic structure. Various first step researches indicate that the double bottom tetraquark states might be "stable" and pending for experimental confirmation. Most of theoretical calculations predict the double bottom tetraquark states, at least $1^+$ state, lie below the open bottom threshold. Whereas for charm quark sector, the corresponding tetraquarks could be loosely bound, above the open charm threshold, or hardly maintain to be resonances.

In this work, we explore the $q q^\prime \bar{Q} \bar{Q}$ tetraquarks with a novel structure, the color-octet-octet tetraquark configuration previously unconsidered, by virtual of the technique of QCD Sum Rules, where $q=u$, $q^\prime = d, s$, and $Q = b, c$. Moreover, we calculate also the four-quark molecular state. Our numerical results indicate that the states of the S-wave color-octet-octet $qq^\prime\bar{b}\bar{b}$ are above the relevant open bottom thresholds and hence may decay into bottom mesons, which is different from what given in the literature. Therefore to observe the tetraquark states via double-meson final states is an interesting issue to explore in experiment. We give out the decays widths of these states.

In contrast, we find there exist two molecular states with masses $10.36$ GeV and $10.48$ GeV for the $ud\bar{b}\bar{b}$ and $us\bar{b}\bar{b}$ configurations, respectively, which lie below the open bottom thresholds. For more analysis on the properties of the bounded doubly bottom four states one may refer to \cite{Francis:2016hui}. For charm sector, we find both color-octet-octet and the color-singlet-singlet tetraquark structures are above the corresponding two charm-meson thresholds.

To estimate the calculation uncertainty and reliability, we analyze schematically the mass dependence on radiative corrections and find it may only induce small corrections, -0.043 to 0.161 GeV, to the results.

\vspace{.7cm} {\bf Acknowledgments} \vspace{.3cm}

Authors are highly grateful to Randy Lewis's helpful discussions. This work was supported in part by the Ministry of Science and Technology of the Peoples' Republic of China(2015CB856703); by the Strategic Priority Research Program of the Chinese Academy of Sciences, Grant No.XDB23030100; and by the National Natural Science Foundation of China(NSFC) under the Grants 11375200 and 11605039, Natural Science Foundation of Hebei Province with contract No. A2017205124, and Science Foundation of Hebei Normal University under Contract No. L2016B08.


\begin{widetext}
\appendix

\section{The spectral densities for cases A to D}
For all currents showned in Eqs. (\ref{current-1}- \ref{current-4}),  we obtain the spectral densities as follows:
\begin{eqnarray}
\rho^{pert}_i (s) &=& \frac{1}{\pi^6} \int^{\alpha_{max}}_{\alpha_{min}} d \alpha \int^{1 - \alpha}_{\beta_{min}} d \beta \bigg{\{} \frac{{\cal F}^3_{\alpha \beta} (\alpha + \beta - 1)}{3^2 \times 2^{11} \alpha^3 \beta^3}\bigg [ {\cal F}_{\alpha \beta} (9 \alpha + 9 \beta - 11) \nonumber \\
& -& 8 m_Q^2 \left(\alpha^2 + \alpha (2 \beta - 1) + (\beta - 1) \beta \right) \bigg]- {\cal N}_i \frac{m_q m_Q {\cal F}^2_{\alpha \beta}(\alpha + \beta - 1)}{3^2 \times 2^9 \alpha^2 \beta^3} \nonumber \\
&\times& \big[ {\cal F}_{\alpha \beta} (7 \alpha + 7 \beta - 9 ) - 6 m_Q^2 ( \alpha^2 + \alpha  (2 \beta - 1) + (\beta - 1)\beta )\big] \nonumber \\
&+& {\cal N}_i \frac{m_{q^\prime} m_Q {\cal F}^3_{\alpha \beta}(\alpha + \beta - 1)}{3 \times 2^8 \alpha^3 \beta^2} - \frac{m_q m_{q^\prime} m_Q^2 {\cal F}^2_{\alpha \beta} (\alpha + \beta - 1)}{2^8 \alpha^2 \beta^2} \bigg{\}} \; , \label{A-1} \\
\rho_i^{\langle \bar{q} q \rangle}(s) &=& \frac{\langle \bar{q} q \rangle}{\pi^4}
\int^{\alpha_{max}}_{\alpha_{min}} d \alpha \bigg{\{}
\int^{1 - \alpha}_{\beta_{min}} d \beta \bigg[ {\cal N}_i
\frac{ m_Q {\cal F}_{\alpha \beta }}{3^2 \times 2^6 \alpha \beta^2} \big[ {\cal F}_{\alpha \beta} (5 \alpha + 5 \beta - 6)\nonumber\\
& -& 4 m_Q^2 \left(\alpha^2 + \alpha (2 \beta - 1) + (\beta - 1) \beta \right)\big] + \frac{ m_q {\cal F}_{\alpha \beta} }{3^2 \times 2^7 \alpha \beta} \big[ 5 {\cal F}_{\alpha \beta} - 4 m_Q^2 (\alpha + \beta)\big]\nonumber \\
 &+& \frac{ m_{q^\prime} m_Q^2 {\cal F}_{\alpha \beta}}{3 \times 2^5 \alpha \beta } \bigg]-
[ \frac{ m_{q} {\cal H}_\alpha^2 }{3^2 \times 2^7  (\alpha - 1) \alpha} +  {\cal N}_i \frac{ m_{q} m_{q^\prime} m_Q {\cal H}_\alpha }{3 \times 2^6 \alpha}] \bigg{\}}\; , \\
\rho_i^{\langle \bar{q^\prime} q^\prime \rangle} (s) &=& \frac{\langle \bar{q^\prime} q^{\prime} \rangle}{\pi^4} \int^{\alpha_{max}}_{\alpha_{min}} d \alpha \bigg{\{} \int^{1 - \alpha}_{\beta_{min}} d \beta \bigg[ {\cal N}_i \frac{m_q m_{q^\prime}}{3^2 \times 2^6 \beta} \left( 2 m_Q^3 (\alpha + \beta) - 3 m_Q {\cal F}_{\alpha \beta}\right)\nonumber\\
&-& {\cal N}_i \frac{m_Q {\cal F}^2_{\alpha \beta}}{3 \times 2^6 \alpha^2 \beta} + \frac{m_q m_Q^2 {\cal F}_{\alpha \beta}}{3 \times 2^5 \alpha \beta} + \frac{m_{q^\prime} {\cal F}_{\alpha \beta}}{3^2 \times 2^7 \alpha \beta}\left(5 {\cal F}_{\alpha \beta} - 4 m_Q^2 (\alpha + \beta)\right)\bigg]  \nonumber \\
&+&   \bigg[ - \frac{m_{q^\prime} {\cal H}_\alpha^2}{3^2 \times 2^7 (\alpha - 1) \alpha } + {\cal N}_i \frac{m_q m_{q^\prime} m_Q {\cal H}_\alpha}{3^2 \times 2^6 (\alpha - 1)}\bigg]\bigg{\}} \; , \\
\rho_i^{\langle G^2 \rangle}(s) &=& \frac{\langle g_s^2 G^2\rangle}{\pi^6} \int^{\alpha_{max}}_{\alpha_{min}} d \alpha \int^{1 - \alpha}_{\beta_{min}} d \beta \bigg{\{} \frac{{\cal F}_{\alpha \beta}}{3^3 \times 2^{13} \alpha^2 \beta^2} \bigg[ {\cal F}_{\alpha \beta} ( 15 \alpha^2 + 2 \alpha (10 \beta - 9) \nonumber\\
&+& 5 \beta^2) - 4 m_Q^2 \left( 3 \alpha^3 + \alpha^2 (7 \beta - 3) + \alpha \beta (5 \beta - 4) + (\beta - 1) \beta^2 \right)\bigg] \nonumber \\
&-& {\cal N}_i \frac{m_q m_Q}{3^3 \times 2^{12} \alpha \beta}\bigg[ {\cal F}_{\alpha \beta} (3 \alpha +3 \beta + 2) - 2 m_Q^2 ( \alpha^2 + \alpha (2 \beta - 1)\nonumber \\
 &+& (\beta - 1) \beta)\bigg] +  {\cal N}_i \frac{m_{q^\prime} m_Q {\cal F}_{\alpha \beta}}{3 \times 2^{12} \alpha \beta} \bigg{\}} \; , \\
\rho_i^{\langle \bar{q} G q \rangle}(s) &=& \frac{\langle g_s \bar{q} \sigma \cdot G q \rangle}{\pi^4} \int_{\alpha_{min}}^{\alpha_{max}} \bigg{\{} \int_{\beta_{min}}^{1 - \alpha} d \beta \bigg[ {\cal N}_i \frac{m_Q}{5\times 13 \times 3^2 \times 2^3 \beta^2}
\left( {\cal F}_{\alpha \beta} (3 \alpha - 9 \beta -4)\right. \nonumber \\
&+&
\left. 2 m_Q^2 ( - \alpha^2 + 2 \alpha \beta + \alpha + 3\beta^2 + \beta) \right) + \frac{m_q}{3^2 \times 2^{10} \beta} \left( 3 {\cal F}_{\alpha \beta} - 2 m_Q^2 (\alpha + \beta)\right) + \frac{m_{q^\prime} m_Q^2}{3 \times 2^9 \beta} \bigg] \nonumber \\
&+ &
 \bigg[ {\cal N}_i \frac{m_Q {\cal H}_\alpha}{3^2 \times 2^7 (\alpha - 1)} + \frac{m_q}{3^2 \times 2^{10}(\alpha - 1)} \left( {\cal H}_\alpha (16 \alpha - 17) - 8 m_Q^2 (\alpha - 1) \right) \nonumber \\
 &+& \frac{m_{q^\prime} m_Q^2}{3 \times 2^7} + {\cal N}_i \frac{m_q m_{q^\prime} m_Q (16 \alpha - 19)}{3^2 \times 2^{10}}\bigg] \bigg{\}}\; ,\\
\rho_i^{\langle \bar{q^\prime} G q^\prime\rangle}(s) &=& \frac{\langle g_s \bar{q^\prime} \sigma \cdot G q^\prime \rangle}{\pi^4} \int_{\alpha_{min}}^{\alpha_{max}} d \alpha \bigg{\{}\int_{\beta_{min}}^{1 - \alpha} d \beta \bigg[ \frac{m_{q^\prime}}{3^3 \times 2^{10} \alpha} \left(3 {\cal F}_{\alpha \beta} - 2 m_Q^2(\alpha + \beta) \right)  \nonumber \\
&-& {\cal N}_i \frac{m_q m_{q^\prime} m_Q}{3^3 \times 2^{10}}\bigg]+ \bigg[ \frac{m_q m_Q^2}{3 \times 2^7}- {\cal N}_i \frac{m_Q {\cal H}_\alpha}{3 \times 2^7 \alpha} - {\cal N}_i \frac{ m_q m_{q^\prime} m_Q (24 \alpha - 5)}{3^3 \times 2^{10}}\nonumber\\
& + &\frac{m_{q^\prime}}{3^3 \times 2^{10} \alpha} [{\cal H}_\alpha (48 \alpha - 5) - 24 m_Q^2 \alpha]\bigg] \bigg{\}}\; , \\
\rho_i^{\langle \bar{q} q \rangle \langle \bar{q^\prime} q^\prime \rangle}(s) &=& \frac{\langle \bar{q} q \rangle \langle \bar{q^\prime} q^\prime \rangle}{\pi^2} \int_{\alpha_{min}}^{\alpha_{max}} d \alpha \bigg[ \frac{m_Q^2}{3^2 \times 2^3} + {\cal N}_i \frac{m_q m_Q (\alpha - 1)}{3^2 \times 2^3}\nonumber\\
& -& {\cal N}_i \frac{m_{q^\prime} m_Q \alpha}{3^2 \times 2^4} - \frac{m_q m_{q^\prime} (\alpha - 1) \alpha}{3 \times 2^5} \bigg] \; , \\
\rho_i^{\langle G^3 \rangle}(s) &=& \frac{\langle g_s^3 G^3 \rangle}{\pi^6} \int_{\alpha_{min}}^{\alpha_{max}} d\alpha \int_{\beta_{min}}^{1 - \alpha} d \beta \bigg[ - \frac{(\alpha + \beta - 1)}{3^3 \times 2^{13} \alpha^3 \beta^3} \big[ 2 m_Q^2 (\alpha^4 (\beta + 2) + \alpha^3 (\beta - 1) \beta \nonumber\\
&+& \alpha^2 \beta^3 + \alpha (\beta - 1) \beta^3 + 2 \beta^4 )- {\cal F}_{\alpha \beta} (\alpha + \beta) (3 \alpha + 3 \beta - 5) (\alpha^2 - \alpha \beta + \beta^2)\big]\nonumber\\
& -& {\cal N}_i \frac{m_q m_Q}{3^3 \times 2^{13} \alpha^2 \beta^3} (\alpha + \beta - 3)(\alpha + \beta -1) (6 \alpha^3 + \beta^3) \nonumber \\
&+& {\cal N}_i \frac{m_{q^\prime} m_Q}{3^2 \times 2^{12} \alpha^3 \beta^2} (\alpha + \beta - 1) (\alpha^3 + 6 \beta^3)\bigg] \; , \\
\rho_i^{\langle G^2\rangle^2} (s) &=& \frac{\langle g_s^2 G^2 \rangle^2 }{3^3 \times 2^{16} \pi^6} \int_{\alpha_{min}}^{\alpha_{max}} d \alpha \bigg{\{}\int_{0}^{1 - \alpha} d \beta  - 5 \bigg{\}} \; , \\
\Pi_i^{\langle G^2 \rangle}(M_B^2) &=& \frac{\langle g_s^2 G^2 \rangle}{ \pi^6 } \int_{0}^1 d \alpha \int_0^{1 - \alpha} d \beta \; e^{-\frac{m_Q^2 (\alpha + \beta)}{M_B^2 \alpha \beta}} \bigg{\{} - {\cal N}_i \frac{ m_q m_Q^5}{3^3 \times 2^{10}\alpha^3 \beta^4} (\alpha^2 - \alpha \beta + \beta^2)   \nonumber \\
&\times&( \alpha^2 + \alpha (2 \beta - 1)+ (\beta - 1) \beta^2)\nonumber\\
&+& \frac{ m_q m_{q^\prime} m_Q^4}{3^2 \times 2^{10} \alpha^3 \beta^3} \left(\alpha^4 + \alpha^3 (\beta - 1) + \alpha \beta^3 + (\beta - 1) \beta^3 \right)\bigg{\}} \; ,  \\
\Pi_i^{\langle \bar{q} G q\rangle}(M_B^2) &=& - {\cal N}_i \frac{\langle g_s \bar{q} \sigma \cdot G q \rangle m_q m_{q^\prime} m_Q^3}{3^2 \times 2^7 \pi^4}\int_{0}^{1}\frac{d \alpha}{\alpha} \; e^{-\frac{m_Q^2}{M_B^2 (1 - \alpha) \alpha}} \; , \\
\Pi_i^{\langle \bar{q^\prime} G q^\prime \rangle}(M_B^2)&=& \frac{\langle g_s \bar{q^\prime} \sigma \cdot G q^\prime \rangle}{\pi^4} \int_{0}^{1} d \alpha \bigg{\{} {\cal N}_i \int_0^{1 - \alpha} d \beta \; e^{- \frac{m_Q^2 (\alpha + \beta)}{M_B^2 \alpha \beta}} \frac{(\alpha + \beta)}{3^3 \times 2^9 \alpha \beta} \nonumber\\
&+& {\cal N}_i e^{- \frac{m_Q^2}{M_B^2 (1 - \alpha) \alpha}} \frac{m_q m_{q^\prime} m_Q^3}{3^2 \times 2^7 (\alpha - 1)} \bigg{\}} \; ,\\
\Pi_i^{\langle \bar{q} q \rangle \langle \bar{q^\prime} q^\prime \rangle} (M_B^2) &=& \frac{\langle \bar{q} q \rangle \langle \bar{q^\prime} q^\prime \rangle}{\pi^2} \int_{0}^1 d \alpha \; e^{- \frac{m_Q^2}{M_B^2 (1 - \alpha) \alpha}} \bigg[ - {\cal N}_i \frac{m_q m_Q^3}{3^2 \times 2^4 \alpha} + {\cal N}_i \frac{m_{q^\prime} m_Q^3}{ 3^2 \times 2^4 (\alpha  -1 )} \nonumber \\
&-& \frac{m_q m_{q^\prime} m_Q^2}{3^2 \times 2^5 M_B^2 (\alpha - 1) \alpha} \left( m_Q^2 - 3 M_B^2 (\alpha - 1) \alpha \right)\bigg] \; ,\\
\Pi_i^{\langle G^3 \rangle} (M_B^2) &=& \frac{\langle g_s^3 G^3 \rangle}{\pi^6} \int_{0}^{1} d \alpha \int_{0}^{1 - \alpha} d \beta \; e^{- \frac{m_Q^2 (\alpha + \beta)}{M_B^2 \alpha \beta}} \bigg[ \frac{m_Q^4}{3^3 \times 2^{11}\alpha^4 \beta^4} (\alpha + \beta - 1)^2 (\alpha + \beta) (\alpha^4 + \beta^4)\nonumber \\
& +& \frac{m_q m_Q^3 (\alpha + \beta - 1)}{3^3 \times 2^{12} \alpha^4 \beta^5 M_B^2} [2 m_Q^2 (\alpha + \beta - 1) (\alpha + \beta)(\alpha^4 + \beta^4)\nonumber\\
& -& M_B^2 \alpha \beta \left(7 \alpha^5 + \alpha^4 (13 \beta - 5) + 6 \alpha^3 (\beta - 1) \beta + \alpha^2 \beta^3 + \alpha \beta^3 (3 \beta -1) + 2 \beta^5 \right)] \nonumber \\
&-&\frac{m_{q} m_{q^\prime} m_Q^2 }{3^2 \times 2^{11} M_B^2 \alpha^4 \beta^4}(\alpha + \beta - 1) \left(m_Q^2 (\alpha^4 + \beta^4)  - 3 M_B^2 \alpha \beta (\alpha^3 + \beta^3) \right)\nonumber\\
& -&\frac{m_{q^\prime} m_Q^3}{3^2 \times 2^{11}\alpha^4 \beta^3} (\alpha + \beta - 1) (\alpha^4 + \beta^4) \bigg] \; , \\
\Pi_i^{\langle \bar{q} q\rangle \langle \bar{q^\prime} G q^\prime \rangle}(M_B^2) &=& \frac{\langle \bar{q} q\rangle \langle g_s \bar{q^\prime} \sigma \cdot G q^\prime \rangle}{\pi^2} \int_0^1 d \alpha \; e^{- \frac{m_Q^2}{M_B^2 (\alpha - 1) \alpha}} \bigg[ \frac{m_Q^2}{3^2 \times 2^5 M_B^2 (\alpha - 1) \alpha} \left( m_Q^2 - M_B^2 (\alpha - 1) \alpha \right)\nonumber \\
&-& {\cal N}_i \frac{m_q m_Q}{3^2 \times 2^6 M_B^4 (\alpha - 1) \alpha^2} \left( 2 M_B^4 (\alpha - 1)^2 \alpha^2 - 2 M_B^2 m_Q^2 (\alpha - 1) \alpha + m_Q^4 \right) \nonumber \\
&+& {\cal N}_i \frac{m_{q^\prime} m_Q}{3^3 \times 2^8 M_B^4 (\alpha - 1)^2 \alpha}[- 2 M_B^4 (\alpha - 1)^2 \alpha + M_B^2 m_Q^2 \left( (9 - 8 \alpha) \alpha - 1) + 8 m_Q^4\right)] \nonumber \\
& +& \frac{m_q m_{q^\prime}}{3^3 \times 2^9 M_B^6 (\alpha - 1)^2 \alpha^2} ( - 4 M_B^6 (\alpha - 1)^3 \alpha^2 - M_B^4 m_Q^2 (\alpha - 1)^2 \alpha (16 \alpha - 3) \nonumber\\
&+& M_B^2 m_Q^4 (\alpha - 1)(16 \alpha - 1) - 8 m_Q^6 )\bigg] \; , \\
\Pi_i^{\langle \bar{q^\prime} q^\prime \rangle \langle \bar{q} G q\rangle}(M_B^2) &=& \frac{\langle \bar{q^\prime} q^\prime \rangle \langle g_s \bar{q} \sigma \cdot G q\rangle}{\pi^2} \int_0^1 d \alpha \; e^{- \frac{m_Q^2}{M_B^2 (1- \alpha) \alpha}} \bigg[ \frac{m_Q^2}{3^2 \times 2^7 M_B^2 (\alpha - 1) \alpha} \left( M_B^2 (5 - 4 \alpha) \alpha + 4 m_Q^2 \right) \nonumber \\
&-& {\cal N}_i \frac{m_q m_Q}{3^3 \times 2^8 M_B^4 (\alpha - 1) \alpha^2} [ M_B^4 (\alpha - 1) \alpha^2 (16 \alpha - 19) + M_B^2 m_Q^2 (19 - 16 \alpha)\alpha + 8 m_Q^4 ] \nonumber \\
&-& \frac{m_q m_{q^\prime} m_Q^2}{3^3 \times 2^9 M_B^6 (\alpha - 1)^2 \alpha^2 } [M_B^4 (\alpha - 1) \alpha^2 (16 \alpha - 19) + M_B^2 m_Q^2 (19 - 16 \alpha) \alpha +8 m_Q^4] \nonumber \\
&+& {\cal N}_i \frac{m_{q^\prime} m_Q^3}{3^2 \times 2^8 M_B^4 (\alpha - 1)^2 \alpha}\left( M_B^2 (5 - 4 \alpha) \alpha + 4 m_Q^2 \right)\bigg] \; , \\
\Pi_i^{\langle G^2 \rangle^2}(M_B^2) &=& \frac{\langle g_s^2 G^2 \rangle^2}{\pi^6} \int_{0}^1 d \alpha \bigg{\{} \int_{0}^{1 - \alpha} \beta \; e^{- \frac{m_Q^2 (\alpha + \beta)}{M_B^2 \alpha \beta}} \bigg[ \frac{m_Q^2}{3^5 \times 2^{15} M_B^4 \alpha^3 \beta^3} [9 M_B^4 \alpha^2 \beta^2 (\alpha + \beta)  \nonumber \\
&-& 2 M_B^2 m_Q^2 \alpha \beta (\alpha + \beta - 1) (3 \alpha + 3\beta - 1)+4 m_Q^4 (\alpha + \beta - 1)^2 (\alpha + \beta)]\nonumber\\
&+&   {\cal N}_i \frac{m_q m_Q^3 (\alpha + \beta - 1)}{3^5 \times 2^{14} M_B^6 \alpha^3 \beta^4} [3 M_B^4 \alpha \beta^2 (3 \alpha + 3 \beta - 1) + M_B^2 m_Q^2 \beta (-11 \alpha^2 + \alpha (9 - 17 \beta) \nonumber \\
&-& 6(\beta - 1) \beta) + 2 m_Q^4 (\alpha + \beta - 1)(\alpha + \beta)] -  {\cal N}_i \frac{m_{q^\prime} m_Q^3 (\alpha + \beta - 1)}{3^4 \times 2^{13} M_B^4 \alpha^3 \beta^2} (m_Q^2 - 3 M_B^2 \alpha)\nonumber \\
&-& \frac{m_q m_{q^\prime} m_Q^2 (\alpha + \beta - 1)}{3^4 \times 2^{13} M_B^6 \alpha^3 \beta^3} (m_Q^2 - 3 M_B^2 \alpha)(m_Q^2 - 3 M_B^2 \beta) \bigg]  \nonumber \\
&-& e^{- \frac{m_Q^2}{ M_B^2 (1- \alpha) \alpha}} \pi^2 \bigg[  \frac{m_{q^\prime} m_Q^2}{3^4 \times 2^{15} M_B^6 (\alpha - 1) \alpha^3} (2 M_B^2 (\alpha - 1) \alpha + m_Q^2) \nonumber\\
&+&{\cal N}_i \frac{m_Q (m_Q^2 - 3 M_B^2 \alpha)}{3^4 \times 2^{14} M_B^4 \alpha^3} \bigg] \bigg{\}}\; , \label{A-17}
\end{eqnarray}
where $M_B$ is the Borel parameter introduced by the Borel
transformation, $q = u, \, q^\prime = d$ or $s$, $Q = c$ or $b$, the subscript $i$ runs from $A$ to $D$, and the factor ${\cal N}_i$ has the following definition: ${\cal N}_A = {\cal N}_B =1$ and ${\cal N}_C = {\cal N}_D = -1$. It should be noted that when $i = B$ ( $i = D$ ), the the letters $q$ and $q^\prime$ in Eqs.(\ref{A-1}-\ref{A-17}) have to be exchanged, since there exist a symmetry between the currents expressed in Eq. (\ref{current-1}) and  Eq. (\ref{current-2}) ( Eq. (\ref{current-3}) and Eq.(\ref{current-4})). Here, we also have the following definitions:
\begin{eqnarray}
{\cal F}_{\alpha \beta} &=& (\alpha + \beta) m_Q^2 - \alpha \beta s \; , {\cal H}_\alpha  = m_Q^2 - \alpha (1 - \alpha) s \; , \\
\alpha_{min} &=& \left(1 - \sqrt{1 - 4 m_Q^2/s} \right) / 2, \; , \alpha_{max} = \left(1 + \sqrt{1 - 4 m_Q^2 / s} \right) / 2  \; , \\
\beta_{min} &=& \alpha m_Q^2 /(s \alpha - m_Q^2).
\end{eqnarray}

\section{The spectral densities for case $E$}

\begin{eqnarray}
\rho^{pert}(s) &=& \frac{1}{\pi^6} \int^{\alpha_{max}}_{\alpha_{min}} d \alpha \int^{1- \alpha}_{\beta_{min}} d \beta  \frac{{\cal F}^2_{\alpha \beta} (\alpha + \beta - 1)}{3^2 \times 2^5 \alpha^3 \beta^3} \nonumber\\
&\times& \bigg{\{}12 m_Q^2 \alpha \bigg[ m_Q m_{q^\prime} \left( \alpha^2 + \beta \left( \beta - 1 \right) + \alpha \left( 2 \beta - 1 \right) \right) \nonumber \\
&+& m_q \left( -6 m_{q^\prime} \beta + m_Q \left( \alpha^2 + \beta \left( \beta - 1 \right) + \alpha \left( 2 \beta - 1 \right) \right) \right)  \bigg]  \nonumber \\
&-& 2 {\cal F}_{\alpha \beta} m_Q \bigg [ 4 m_Q \left( \alpha^2 + \left( \beta -1 \right) \beta + \alpha \left( 2 \beta - 1 \right) \right) + m_q \left( 7 \alpha^2 - 6 \beta + \alpha \left( 7 \beta - 9 \right) \right)    \nonumber \\
&+& m_{q^\prime} \left( 7 \alpha^2 - 6 \beta + \alpha \left( 7 \beta - 9 \right) \right)  \bigg]+  {\cal F}^2_{\alpha \beta} (9 \alpha + 9 \beta -11) \bigg{\}} \; , \\
\rho^{\langle \bar{q} q \rangle}(s) &=& \frac{\langle \bar{q} q \rangle}{\pi^4} \int^{\alpha_{max}}_{\alpha_{min}} d \alpha \bigg{\{} \int^{1- \alpha}_{\beta_{min}} d \beta \bigg[ \frac{{\cal F}^2_{\alpha \beta}}{3^2 \times 2^2 \alpha \beta^2} \big[ 5\beta \left( m_q + m_{q^\prime} \right) +4 m_Q \left( 5 \alpha + 5 \beta \right) \big]  \nonumber \\
&-& \frac{ m_Q^2 {\cal F_{\alpha \beta}} }{3^2 \alpha \beta^2} \beta \left( m_q + m_{q^\prime} \right) \left( \alpha + \beta -3 \right) + \frac{4 m_Q^3 {\cal F}_{\alpha \beta}}{3^2 \alpha \beta^2} \big[ \alpha^2 + \beta^2 - \beta + \alpha \left( 2 \beta - 1 \right) \big]   \bigg] \nonumber \\
&-& \frac{{\cal H}_\alpha}{3^2 \times 2^2 \alpha ( \alpha - 1 )} \bigg[ {\cal H}_\alpha \left( m_q + m_{q^\prime} \right) + 12 m_q m_{q^\prime} m_Q \left( \alpha -1 \right) \bigg] \bigg{\}} \; ,\\
\rho^{\langle \bar{q^\prime} q^\prime \rangle} (s) &=& \frac{\langle \bar{q^\prime} q^\prime \rangle}{\pi^4} \int^{\alpha_{max}}_{\alpha_{min}} d \alpha \bigg{\{} \int^{1 - \alpha}_{\beta_{min}} d \beta \bigg[ \frac{{\cal F}^2_{\alpha \beta}}{3^2 \times 2^2 \alpha^2 \beta} \big[ -12 m_Q + 5 \left( m_q + m_{q^\prime} \right) \big] \nonumber \\
 &+&\frac{2 m_q m_{q^\prime} m_Q^3 \alpha^2 (\alpha + \beta - 3)}{3^2 \alpha^2 \beta}+ \frac{{\cal F}_{\alpha \beta} m_Q \alpha}{3^2 \alpha \beta} \big[ m_Q \left( m_q + m_{q^\prime} \right)  \left( \alpha + \beta -3 \right) + 3 m_q  m_{q^\prime} \alpha \big] \bigg] \nonumber \\
&+& \frac{{\cal H}_\alpha}{3^2 \times 2^2 ( \alpha - 1 ) \alpha} \bigg[ 4 m_q m_{q^\prime} m_Q - {\cal H}_\alpha \left( m_q + m_{q^\prime} \right) \bigg] \bigg{\}} \; ,\nonumber \\
\rho^{\langle G^2 \rangle}(s) &=& \frac{\langle g_s^2 G^2 \rangle}{\pi^6} \int^{\alpha_{max}}_{\alpha_{min}} d \alpha \int^{1-\alpha}_{\beta_{min}} d \beta \bigg{\{} \frac{1}{3^2 \times 2^8 \alpha^2 \beta^2} \bigg[ -2 m_Q^3 \left( m_q + m_{q^\prime} \right) \alpha \beta \nonumber\\
&\times&\left( \alpha + \beta - 1 \right) \left( \alpha + \beta \right) - {\cal F}_{\alpha \beta}^2 \left( 15 \alpha^2 + 20 \alpha \beta - 18 \alpha +5 \beta^2 \right) \nonumber \\
&+& {\cal F}_{\alpha \beta} m_Q \left( \left( \alpha + \beta - 1 \right) \left( \alpha + \beta \right) \left( 3 \alpha + \beta \right) + \left( m_q + m{q^\prime} \right) \alpha \beta \left( 3 \alpha + 3 \beta -7 \right) \right) \bigg] \nonumber \\
&-& \frac{m_Q (\alpha + \beta - 1)}{3^3 \times 2^6 \alpha^3 \beta^3} \bigg[ {\cal F}_{\alpha \beta} \big[ \left( 3 m_q + 3 m_{q^\prime} \right) \left( 3 \alpha^4 -6 \beta^3 + 3 \alpha^3 \beta - 5 \alpha^3 \right)\nonumber\\
& -& 2 m_Q \left( 3 \alpha^4 - 5 \alpha^3 + 3 \alpha^3 \beta + 3 \alpha \beta^3 + 3 \beta^4 - 5 \beta^3 \right) \big] \nonumber \\
&+& m_Q \big[ m_q ( 36 m_{q^\prime} ( \alpha^3 \beta + \alpha \beta^3 ) + m_Q ( -5 \alpha^5 + 3 \alpha^4 - 11 \alpha^4 \beta - 6 \alpha^3 \beta^2  \nonumber \\
&+& \alpha^2 \beta^3 - \beta^3 + \beta^4 ) ) + m_Q \left( \alpha + \beta \right) ( 4 m_Q \left( \alpha^4 + \alpha^3 \beta - \alpha^3 + \alpha \beta^3 - \beta^3 + \beta^4 \right)\nonumber\\
& +& m_{q^\prime} \left( -5 \alpha^4 -6 \alpha^3 \beta + \alpha^3 - 3 \alpha^2 \beta - 6 \beta^3 + 3 \alpha \beta^2 + \alpha \beta^3 \right) ) \big] \bigg] \bigg{\}} \; ,\\
\rho^{\langle \bar{q} G q \rangle}(s) &=& \frac{\langle g_s \bar{q} \sigma \cdot G q \rangle}{\pi^4} \int_{\alpha_{min}}^{\alpha_{max}} \bigg{\{} \int_{\beta_{min}}^{1-\alpha} d \beta \bigg[ \frac{1}{3^2 \times 2 \beta} \left( -3 m_Q {\cal F}_{\alpha \beta} + 2 m_Q^3 (\alpha + \beta) \right) \nonumber \\
&+& \frac{1}{3 \times 2^5 \beta^2} \big[ - {\cal F}_{\alpha \beta } \left( 3 \left( m_q + m_{q^\prime} \right) +4 m_Q \left( 3 \alpha + 3 \beta -4 \right) \right) \nonumber \\
&+& 2 m_Q^2 \left( \beta \left( m_q + m_{q^\prime} \right) \left( \alpha + \beta -3 \right) + 4m_Q \left( \alpha + \beta - 1 \right) \left( \alpha + \beta \right) \right)  \big] \bigg] \nonumber \\
&+& \frac{1}{3^2 \times 2 (\alpha - 1)} \bigg[ {\cal H}_\alpha \left( m_Q + \left( m_q + m_{q^\prime} \right) \left( \alpha - 1 \right) \right) + m_Q \left( \alpha - 1 \right) ( m_Q \left( m_q + m_{q^\prime} \right)\nonumber\\
& +& 2 m_q  m_{q^\prime} \left( \alpha -1 \right) ) \bigg] + \frac{1}{3 \times 2^5 (\alpha - 1)} \big[ {\cal H}_\alpha \left( m_q + m_{q^\prime} \right) + 6 m_q m_{q^\prime} m_Q \left( \alpha - 1 \right) \big] \bigg{\}} \;, \\
\rho^{\langle \bar{q^\prime} G q^\prime \rangle}(s) &=& \frac{\langle g_s \bar{q^\prime} \sigma \cdot G q^\prime \rangle}{\pi^4} \int_{\alpha_{min}}^{\alpha_{max}} d \alpha \bigg{\{} \int_{\beta_{min}}^{1-\alpha} d \beta  \frac{1}{3^2 \times 2^5 \alpha}\bigg[  -3 {\cal F}_{\alpha \beta} \left( m_q + m_{q^\prime} \right) \nonumber \\
&+& 2 m_Q^2 \left( m_q + m_{q^\prime} \right) \left( \alpha + \beta \right) + 2 \alpha m_Q m_q m_{q^\prime}  \bigg]\nonumber\\
& +&  \frac{1}{3 \times 2^5 \alpha} \bigg[ {\cal H}_\alpha \left( m_q + m_q^\prime \right) -2 \alpha m_Q m_q m_{q^\prime} \bigg]\nonumber \\
&+&\frac{1}{3^2 \times 2 \alpha} \bigg[ {\cal H}_{\alpha} \left( -3 m_Q + \alpha m_q + \alpha  m_{q^\prime}  \right) + m_Q \alpha \left( \left( m_q + m_{q^\prime} \right) m_Q  - m_q m_{q^\prime} \alpha \right) \bigg]  \bigg{\}}  \; ,\\
\rho^{\langle \bar{q} q \rangle \langle \bar{q^\prime} q^\prime \rangle}(s) &=& \frac{\langle \bar{q} q \rangle \langle \bar{q^\prime} q^\prime \rangle}{3^3 \pi^2} \int_{\alpha_{min}}^{\alpha_{max}} d \alpha \bigg[ 4 m_Q^2 + m_Q \left( m_q + m_{q^\prime} \right) \left( \alpha - 2 \right) - 3 m_Q m_{q^\prime} \alpha \left( \alpha - 1 \right) \bigg] \; , \\
\rho^{\langle G^3 \rangle}(s) &=& \frac{\langle g_s^3 G^3 \rangle}{\pi^6} \int_{\alpha_{min}}^{\alpha_{max}} d\alpha \int_{\beta_{min}}^{1 - \alpha} d \beta \bigg{\{} \frac{ \alpha + \beta - 1 }{3^3 \times 2^7 \beta^3} \bigg[ {\cal F}_{\alpha \beta} \left( 3 \alpha + 3 \beta - 5 \right) \nonumber \\
&-& m_Q \left( 3 \left( m_q + m_{q^\prime} \right) \left( \alpha^2 + \alpha \beta - 3 \alpha - \beta \right) +  2 m_Q \left( 2 \alpha + \alpha \beta - \beta + \beta^2 \right) \right) \bigg] \nonumber \\
&+& \frac{ \alpha + \beta - 1 }{3^3 \times 2^8 \beta^3} \bigg[ 2 {\cal F}_{\alpha \beta} \left( 3 \alpha + 3 \beta - 5 \right) - m_Q ( \left( m_q + m_{q^\prime} \right) \left( \alpha^2 + \alpha \beta - 3 \alpha - 36 \beta \right) \nonumber\\
&+& 4 m_Q \left( 2 \beta + \alpha \beta - \alpha + \alpha^2 \right) ) \bigg] \bigg{\}} \; ,\\
\Pi^{\langle G^2 \rangle}(M_B^2) &=& \frac{\langle g_s^2 G^2 \rangle}{ \pi^6 } \int_{0}^1 d \alpha \int_0^{1 - \alpha} d \beta \;e^{-\frac{m_Q^2 (\alpha + \beta)}{M_B^2 \alpha \beta}} \bigg{\{} \frac{m_Q^4}{3^3 \times 2^5 \alpha^3 \beta^4} \left( \alpha^4 + \alpha^3 \beta - \alpha^3 +\alpha \beta^3 - \beta^3 + \beta^4 \right) \nonumber \\
&\times& \bigg[ m_Q \left( m_q + m_{q^\prime} \right) \left( \alpha + \beta - 1 \right) \left( \alpha + \beta \right) - 6 \beta m_q m_{q^\prime} \bigg] \bigg{\}} \; ,\\
\Pi^{\langle \bar{q} G q\rangle}(M_B^2) &=& -\frac{\langle g_s \bar{q} \sigma \cdot G q \rangle m_q m_{q^\prime} m_Q^3}{3^2 \times 2 \pi^4}\int_{0}^{1}\frac{d \alpha}{\alpha} \; e^{-\frac{m_Q^2}{M_B^2 (1 - \alpha) \alpha}} \; ,\\
\Pi^{\langle \bar{q^\prime} G q^\prime \rangle}(M_B^2) &=& -\frac{\langle g_s \bar{q^\prime} \sigma \cdot G q^\prime \rangle m_q m_{q^\prime} m_Q^3}{\pi^4}\int_{0}^{1} d \alpha \bigg{\{} \int_0^{1 - \alpha} d \beta \bigg[ \frac{\alpha + \beta}{3^2 \times 2 ^5 \alpha}  e^{- \frac{m_Q^2 (\alpha + \beta)}{M_B^2 \alpha \beta}} \;  \bigg]\nonumber\\
& +& \frac{1}{3^2 \times 2 ( 1 - \alpha)} \;  e^{-\frac{m_Q^2}{M_B^2 (1 - \alpha) \alpha}} \bigg{\}}  \; , \\
\Pi^{\langle \bar{q} q \rangle \langle \bar{q^\prime} q^\prime \rangle} (M_B^2) &=& \frac{\langle \bar{q} q \rangle \langle \bar{q^\prime} q^\prime \rangle}{3^2 \times \pi^2} \int_{0}^1 d \alpha \; e^{- \frac{m_Q^2}{M_B^2 (1 - \alpha) \alpha}} \frac{2 m_q m_{q^\prime} m_Q^2}{M_B^2 (1 - \alpha) \alpha} \bigg[  - M_B^2 m_Q \left( m_q + m_{q^\prime} \right)\nonumber\\
&+&m_q m_{q^\prime} m_Q^2 + 3 M_B^2 m_q m_{q^\prime} \alpha \left( \alpha - 1 \right) \bigg] \; ,\\
\Pi^{\langle G^3 \rangle} (M_B^2) &=& \frac{\langle g_s^3 G^3 \rangle}{\pi^6} \int_{0}^{1} d \alpha \int_{0}^{1 - \alpha} d \beta \; e^{- \frac{m_Q^2(\alpha + \beta)}{M_B^2 \alpha \beta}} \bigg[ \left( \frac{1}{\beta^5} - \frac{1}{\alpha^4 \beta}   \right) \big[ \left( 2 m_Q^3 \left( m_q + m_{q^\prime} \right) + 4 M_B^2 m_Q^2 \beta \right)\nonumber \\
&\times&\left( \alpha + \beta - 1 \right) \left( \alpha + \beta \right) + \beta M_B^2 m_Q \left( m_q + m_{q^\prime} \right) \left( 7 \alpha^2 + 6 \beta^2 + 13 \alpha \beta -5 \alpha \right)\nonumber\\
& -& 12 m_q m_{q^\prime} m_Q^2 \beta + 36 M_B^2 m_q m_{q^\prime} \beta^2 \big] \bigg] \; ,\\
\Pi^{\langle \bar{q} q\rangle \langle \bar{q^\prime} G q^\prime \rangle}(M_B^2) &=& \frac{\langle \bar{q} q\rangle \langle g_s \bar{q^\prime} \sigma \cdot G q^\prime \rangle}{\pi^2} \int_0^1 d \alpha \; e^{- \frac{m_Q^2}{M_B^2 (\alpha - 1) \alpha}} \bigg[ - \frac{m_q m_{q^\prime} m_Q^6}{3^3 M_B^6 {(\alpha - 1)}^2 \alpha^2 } \nonumber \\
&-& \frac{m_Q^3 ( m_q + m_{q^\prime})}{ 3^2 \times 2^3 M_B^2 (\alpha - 1) \alpha} - \frac{m_Q^5 (m_q + m_{q^\prime}) (\alpha - 3)}{3^3 \times 2 M_B^4 {(\alpha - 1)}^2 \alpha^2 } + \frac{2 m_q m_{q^\prime} m_Q^4}{3^3 M_B^4 \alpha (\alpha - 1)} \nonumber \\
&+& \frac{m_Q^2}{3^3 M_B^2 \alpha (\alpha - 1)} \left( - 2 m_q m_{q^\prime} \left( \alpha - 1 \right ) \alpha + m_Q \left( m_q + m_{q^\prime} \right) \left( 2 \alpha - 3 \right) + 6 m_Q^2 \right) \nonumber \\
&+& \frac{m_Q ( m_q + m_{q^\prime})}{3^2 \times 2^2} + \frac{m_Q^4 m_q m_{q^\prime}}{3^2 \times 2^3 m_B^4 (\alpha - 1) \alpha} - \frac{m_Q^2 m_q m_{q^\prime}}{3 \times 2^3 M_B^3 \alpha} + \frac{m_q m_{q^\prime} (\alpha - 1)}{3^2 \times 2 }\nonumber\\
&-& \frac{2 m_Q^2 + m_Q (m_q + m_{q^\prime}) (\alpha - 1)}{3^2}\bigg] \; ,\\
\Pi^{\langle \bar{q^\prime} q^\prime \rangle \langle \bar{q} G q\rangle}(M_B^2) &=& \frac{\langle \bar{q^\prime} q^\prime \rangle \langle g_s \bar{q} \sigma \cdot G q\rangle}{\pi^2} \int_0^1 d \alpha \; e^{- \frac{m_Q^2}{M_B^2 (1- \alpha) \alpha}} \bigg[ - \frac{m_Q^6 m_q m_{q^\prime}}{3^3 M_B^6 {(\alpha - 1)}^2 \alpha^2} \nonumber\\
&-&\frac{2 m_Q}{3^3} \bigg[ \frac{m_Q}{3^3 \times 2 M_B^2 (\alpha - 1) \alpha} \big[ 12 m_Q^2 + m_Q \left( m_q + m_{q^\prime} \right) - 4 m_Q m_q m_{q^\prime} \alpha \left( \alpha - 1 \right) \big] \nonumber \\
&+& \frac{m_Q^3}{3^3 \times 2 M_B^4} \big[ m_Q \left( m_q + m_{q^\prime} \right) \left( \alpha + 2 \right) + 4 m_q m_{q^\prime} \alpha \left( \alpha - 1 \right) \big] \nonumber \\
&+& \frac{m_q m_{q^\prime} m_Q^4}{3 \times 2^3 M_B^4 \alpha^2 {(\alpha - 1)}^2} - \frac{m_Q}{3^2 \times 2 (\alpha - 1)} \big[ 4 m_Q + \left( m_q + m_{q^\prime} \right) \left( \alpha - 1 \right) \big] \nonumber \\
&+& \frac{m_Q^2}{3 \times 2^3 M_B^2 \alpha {(\alpha - 1)}^2} \big[ m_Q \left( m_q + m_{q^\prime} \right) + m_q m_{q^\prime} \alpha \left( \alpha - 1 \right) \big]\nonumber\\
&+&  3 m_Q + \left( m_q + m_{q^\prime} \right) \left( \alpha - 1 \right)\bigg] \; .
\end{eqnarray}
\end{widetext}

\end{document}